\documentclass[preprint,aps,nofootinbib,a4paper,amsmath,amssymb,floatfix]{revtex4}
\usepackage{color}
\usepackage{graphicx}
\usepackage{dcolumn}
\usepackage{bm}
\usepackage{amsmath,amsfonts,amssymb}
\usepackage{bbm}
\usepackage{latexsym}
\usepackage{epsfig}
\usepackage{pbox}
\usepackage{multirow}
\usepackage{rotating}
\usepackage{bm}

\newcommand{\be}{\begin{equation}}
\newcommand{\ee}{\end{equation}}
\newcommand{\beqq}{\setlength\arraycolsep{2pt}\begin{eqnarray}}
\newcommand{\eeqq}{\vspace{0cm} \end{eqnarray}}
\newcommand{\bea}{\begin{eqnarray}}
\newcommand{\eea}{\end{eqnarray}}

\newcommand{\lambdab}{\stackrel{\neg}{\lambda}}

\newcommand{\e}{\textrm{e}}

\begin{document}

\title{Creation of Elko particles in asymptotically expanding universe}

\author{S. H. Pereira} \email{shpereira@feg.unesp.br}
\author{Rodrigo C. Lima} \email{castro.lima.rodrigo@gmail.com}

\affiliation{Universidade Estadual Paulista (Unesp)\\Faculdade de Engenharia, Guaratinguet\'a \\ Departamento de F\'isica e Qu\'imica\\ Av. Dr. Ariberto Pereira da Cunha 333 - Pedregulho\\
12516-410 -- Guaratinguet\'a, SP, Brazil}




\begin{abstract}
In the present article we study the process of particle creation for Elko spinor fields as a consequence of expansion of the universe. We study the effect driven by a expanding background that is asymptotically minkowskian in the past and future. The differential equation that governs the time mode function is obtained for the conformal coupling case and, although its solution is non-analytic, within an approximation that preserves the characteristics of the terms that break the analyticity, analytic solutions are obtained. Thus, by means of the Bogolyubov transformations technique, the number density of particles created is obtained, which can be compared to exact solutions already present in literature for scalar and Dirac particles. The spectrum of created particles is obtained and it was found that it is a generalization of the scalar field case, which converges to the scalar field one when the specific terms concerning the Elko field are dropped out. We also found that lighter Elko particles are created in larger quantities than Dirac fermionic particles. By considering the Elko particles as candidate to dark matter in the universe, such result shows that there are more light dark matter (Elko) particles created by gravitational effects in the universe than baryonic (fermionic) matter, in agreement to standard model.
\end{abstract}

\maketitle

\section{Introduction}

Particle creation in an expanding universe is one of the most interesting results that follows from quantum field theory when applied to curved backgrounds, and the year of 2016 marks fifty years of Leonard Parker's Ph.D. thesis \cite{parkerThesis,parker} where he developed the techniques to address the problem in a rigorous way. Since then the problem has been studied in several articles \cite{partcreation,staro,pavlov01,pavlov02,gribmama02,fabris01} and books \cite{davies,fulling,grib,mukh2}.

One of the most interesting results that follows from the work of Parker is that there is no creation of massless particles (photons, for instance) in a radiation dominated universe neither super massive particles in a matter dominated universe. This implies that there is no particle creation of the dominant type in the corresponding universe phase.

Two different methods are commonly used in order to study particle production by gravitational fields. The standard method, adopted by Parker and others \cite{davies,fulling,parker}, is the method of the adiabatic vacuum state, where the vacuum state is defined as that one for which the lowest energy state smoothly goes to zero in the past ($t\to 0$) and in the future ($t\to \infty$). The result is equivalent to consider a universe that started from a Minkowski flat space at early times and evolves to a Minkowski flat space at late times. The particle creation occurs between such different stages. Another method is that of instantaneous hamiltonian diagonalization \cite{grib,gribmama02,pavlov01} developed by Grib and Mamayev at the same time of the works of Parker. In such approach the vacuum states are defined as those which minimizes the energy at a particular instant of time. In this method the creation rate is relatively higher than the adiabatic method \cite{pavlov02}. 

Exact solutions for the rate of created particles are a difficult task, depending on the scale factor that describes the evolution of the universe and also on the type of particle that we would consider, namely, fermionic or bosonic particles and massive or massless particles. Particularly, a scale factor of the form $(A+B\tanh(C\eta))^{s}$, with $A,\,B,\,C$ constants and $s=\pm 1,\,\pm 2$ is exactly soluble for scalar particles \cite{duncan,moradi1}, where $\eta$ represents the conformal time. Such kind of scale factor has interesting properties in the limit $A\to 0$ and $0<\eta<1/C$, for which the universe behaves like radiation-dominated with $s=2$ and approaches to de Sitter model with $s=-2$. Also, in the limit $B/A \ll 1$ and $s=1$, this model can be compared to exact solutions for Dirac particles \cite{moradi2}. In this paper we study such case. Another exactly soluble model for scalar particles is given by the scale factor $(1+\exp(C\eta))^{s}$ with $s=\pm 1,\,\pm 2$ \cite{moradi1}, while for $s=-2$ we have solution for Dirac particles \cite{moradi2}.

Other recent works have studied the creation of massive and massless scalar particles \cite{alvarenga,padma,ccdmFS}, including modified gravity theories \cite{saulofR}, super-heavy particles as candidates to dark matter \cite{superDM} and also fermionic spin $1/2$  particles \cite{ghosh} in different backgrounds for different methods. It is important to stress that there are several differences between the calculations of creation of scalar particles and fermionic ones. For some few specific models the calculations are analytic and exactly soluble for both kind of particles. 

Recent works have shown the existence of a new kind of non-standard spinor with interesting properties for the particle physics. It is constructed as a spin-$1/2$ field describing a fermion that is eigenstate of charge conjugation operator (in contrast to Dirac particles, that are eigenstate of parity conjugation operator), and was proposed by Ahluwalia and Grumiller \cite{AHL1,AHL2,AHL3}. Such new type of spinor, in contrast to the Dirac spinor field, will violate either locality or Lorentz invariance, or both. For this reason, sometimes they are called Non-Standard Spinors (NSS), or dark spniors, since that they are naturally neutral, whereas Dirac spinors are charged. A special class of such NSS spinors are named Elko, which has mass dimension one (while Dirac particles has mass dimension $3/2$), which leads them to satisfy only a Klein-Gordon type equation. Moreover, being a neutral field, they are a good candidate to a particle that does not interact electromagnetically with the standard model particles, exactly as desired for dark matter particles. Exact solutions for Elko spinor field coupled to gravity in curved Friedmann-Robertson-Walker backgrounds was investigate in \cite{js}. In this paper we investigate the Elko particle creation using the Parker formalism.

The paper is organized as follows. In Section II we deduce the differential equation that governs the time mode function for the Elko spinor in a general curved background. In Section III we particularize and study the solutions to an asymptotically expanding metric. In Section IV the quantized field is constructed in terms of \emph{in} and \emph{out} solutions for the mode functions and the number density of created particles is obtained. In Section V we compare with the results of scalar and Dirac particles. We conclude in section VI.

\section{General equation for Elko creation}

In which follows we will use $\lambda(x)\equiv\lambda(\vec{x},t)$ for a general Elko spinor defined in a curved background in four dimensions, $\Lambda(\vec{x},\eta)$ for the Elko spinor in the conformal time $\eta$, $\lambda^{S/A}_\beta({\bf k})$ for the Elko spinor basis in flat space-time (see Appendix A), where $S/A$ stands for self-conjugate$/$anti-self-conjugate, $\beta$ for the helicities of the field and ${\mathfrak{f}(\vec{x},\eta)}$ is the quantum field constructed with the above basis. The corresponding duals will be denoted by $\lambdab$ and similarly for the other quantities.

We start with the Elko spinor action in a curved background \cite{js}, 
\begin{equation}\label{pe1}
 S = \int\sqrt{-g}\left(\frac{1}{2}g^{\mu\nu}\nabla_{\mu}\stackrel{\neg}{\lambda}\nabla_{\nu}\lambda - V(\stackrel{\neg}{\lambda},\lambda) - \frac{\xi}{2} R\stackrel{\neg}{\lambda}\lambda\right)d^{4}x,
\end{equation}
where $\lambda(x)$ and $\stackrel{\neg}{\lambda}(x)$ represents the Elko spinor field ant its dual, respectively, which are  coupled to gravity through the coupling $\xi$, being $\xi=0$ a minimal coupling and $\xi=1/6$ a conformal coupling. $R$ is the Ricci scalar, ${V(\stackrel{\neg}{\lambda},\lambda)}$ is the potential, ${g \equiv det (g_{\mu\nu})}$ and the covariant derivatives acting into the Elko fields ${\nabla_{\mu}\stackrel{\neg}{\lambda} = \partial_{\mu}\stackrel{\neg}{\lambda} + \stackrel{\neg}{\lambda}\Gamma_{\mu}}$ and ${\nabla_{\mu}\lambda = \partial_{\mu}\lambda - \Gamma_{\mu}\lambda}$, 
with ${\Gamma_{\mu}}$ being the spin connection.

For a spatially flat, homogeneous and isotropic Friedmann-Robertson-Walker metric, 
\begin{equation}\label{pe2}
 ds^{2} = dt^{2} - a^{2}(t)(dx^{2} + dy^{2} + dz^{2}),
\end{equation}
the standard Dirac matrices $\gamma_\mu$ in a flat space time satisfies $\gamma_\mu\gamma_\nu+\gamma_\nu\gamma_\mu = 2\eta_{\mu\nu}$, with $\eta_{\mu\nu}=diag (1,-1,-1,-1)$, and are related to Dirac matrices in the curved metric (\ref{pe2}) by 
\begin{equation}
\gamma^0(t)=\gamma_{0}\,, \hspace{1cm} \gamma^i(t)=-{1\over a(t)}\gamma_i\,,\quad i=1,\,2,\,3.
\end{equation}
With this, the spin connections are ${\Gamma_{0} = \Gamma^{0}=0}$, ${\Gamma_{i} = -\frac{\dot{a}}{2}\gamma_{0}\gamma_{i}}$ and ${\Gamma^{i} = g^{ij}\Gamma_j=\frac{\dot{a}}{2a^2}\gamma_{0}\gamma_{i}}$. By assuming a potential of the form $V={1\over 2}m^2\stackrel{\neg}{\lambda}\lambda$ and taking the least action principle from the action (\ref{pe1}) with respect to $\stackrel{\neg}{\lambda}$ and $\lambda$ we obtain:
\begin{eqnarray}\nonumber
 \partial_{\alpha}[\sqrt{-g}g^{\alpha\nu}(\partial_{\nu}\lambda - \Gamma_{\nu}\lambda)] &+& \sqrt{-g}[g^{\mu\nu}(\Gamma_{\mu}\Gamma_{\nu}\lambda - \Gamma_{\mu}\partial_{\nu}\lambda) + m^{2}\lambda + \xi R\lambda] = 0,\nonumber \\
 \partial_{\alpha}[\sqrt{-g}g^{\alpha\nu}(\partial_{\nu}\stackrel{\neg}{\lambda} + \stackrel{\neg}{\lambda}\Gamma_{\nu})] &+& \sqrt{-g}[g^{\mu\nu}(\Gamma_{\mu}\Gamma_{\nu}\stackrel{\neg}{\lambda} + \partial_{\mu}\Gamma_{\nu}\stackrel{\neg}{\lambda}) + m^{2}\stackrel{\neg}{\lambda} + \xi R\stackrel{\neg}{\lambda}] = 0.
\end{eqnarray}

The motion equations for the Elko spinors are \cite{js}
\begin{eqnarray}\label{pe3}
\ddot{\lambda} + 3\left(\frac{\dot{a}}{a}\right)\dot{\lambda} - \frac{1}{a^{2}}\partial_{i}^{2}\lambda - \frac{3}{4}\left(\frac{\dot{a}}{a}\right)^{2}\lambda + \frac{\dot{a}}{a^{2}}\gamma_{0}\gamma_{i}(\partial_{i}\lambda) + m^{2}\lambda + \xi R\lambda = 0,\nonumber \\
\ddot{\stackrel{\neg}{\lambda}} + 3\left(\frac{\dot{a}}{a}\right)\dot{\stackrel{\neg}{\lambda}} - \frac{1}{a^{2}}\partial_{i}^{2}\stackrel{\neg}{\lambda} - \frac{3}{4}\left(\frac{\dot{a}}{a}\right)^{2}\stackrel{\neg}{\lambda} - \frac{\dot{a}}{a^{2}}(\partial_{i}\stackrel{\neg}{\lambda})\gamma_{0}\gamma_{i} + m^{2}\stackrel{\neg}{\lambda} + \xi R\stackrel{\neg}{\lambda} = 0.
\end{eqnarray}
Now we rewrite the above equations in terms of the conformal time $\eta = \int dt/a(t)$, so that (\ref{pe2}) turns to $ds^2=a^2(\eta)(d\eta^2 - dx^i dx_i)$ and ${\lambda(\vec{x},t)}$ ${\longrightarrow}$ ${\lambda(\vec{x},\eta)}$, ${a(t)}$
${\longrightarrow}$ ${a(\eta)}$. We also rewrite the field ${\lambda(\vec{x},\eta) =  \frac{\Lambda(\vec{x},\eta)}{a(\eta)}}$ and define $f'\equiv df/d \eta$. With this changes and using 
\begin{equation}\nonumber
 R = \frac{6}{a^{2}}(a\ddot{a} + \dot{a}^{2}) = \frac{6}{a^{2}}\frac{a^{''}}{a},
\end{equation}
we obtain for (\ref{pe3}):
\begin{eqnarray}\label{pe8}
 &&\Lambda^{''} - \partial_{i}^{2}\Lambda + \frac{a^{'}}{a}\gamma_{0}\gamma_{i}(\partial_{i}\Lambda) + \left[m^{2}a^{2}- 6\frac{a^{''}}{a}\left(\frac{1}{6}-\xi\right) - \frac{3}{4}\frac{a^{'2}}{a^{2}}\right]\Lambda = 0\nonumber \\
 &&\stackrel{\neg}{\Lambda}^{''} - \partial_{i}^{2}\stackrel{\neg}{\Lambda} - \frac{a^{'}}{a}(\partial_{i}\stackrel{\neg}{\Lambda})\gamma_{0}\gamma_{i} + \left[m^{2}a^{2}- 6\frac{a^{''}}{a}\left(\frac{1}{6}-\xi\right) - \frac{3}{4}\frac{a^{'2}}{a^{2}}\right]\stackrel{\neg}{\Lambda} = 0.
\end{eqnarray}
In the conformal coupling case $\xi=1/6$, we finally obtain
\begin{eqnarray}\label{pe9}
 &&\Lambda^{''} - \partial_{i}^{2}\Lambda + \frac{a^{'}}{a}\gamma_{0}\gamma_{i}(\partial_{i}\Lambda) + \left[m^{2}a^{2} - \frac{3}{4}\frac{a^{'2}}{a^{2}}\right]\Lambda = 0,\nonumber\\
 &&\stackrel{\neg}{\Lambda}^{''} - \partial_{i}^{2}\stackrel{\neg}{\Lambda} - \frac{a^{'}}{a}(\partial_{i}\stackrel{\neg}{\Lambda})\gamma_{0}\gamma_{i} + \left[m^{2}a^{2} - \frac{3}{4}\frac{a^{'2}}{a^{2}}\right]\stackrel{\neg}{\Lambda} = 0.
\end{eqnarray}

From now on we just consider the equation for $\Lambda$. A similar treatment can be done for $\stackrel{\neg}{\Lambda}$. There are four independent Elko fields, namely $\Lambda = \{ \Lambda_{\beta}^{S}, \Lambda_{\beta}^{A}\}$ 
where
\begin{eqnarray}\label{pe11}
 \Lambda_{\beta}^{S} &=& \Lambda_{\{\pm,\mp\}}^{S} = N\lambda_{\{\pm,\mp\}}^{S}({\bf k})g^{(+)}(\eta)e^{i{\bf k\cdot x}},\nonumber\\
 \Lambda_{\beta}^{A} &=& \Lambda_{\{\pm,\mp\}}^{A} = N\lambda_{\{\pm,\mp\}}^{A}({\bf k})g^{(-)}(\eta)e^{-i{\bf k\cdot x}},
\end{eqnarray}
for the self-conjugate $\Lambda_{\beta}^{S}$ and $\Lambda_{\beta}^{A}$ for the anti-self-conjugate spinors, being ${N}$ a normalization constant, $\beta=\{\pm,\mp\}$ the helicities and ${g^{\pm}(\eta)}$ the positive and negative frequencies time functions. Substituting into (\ref{pe11}) we obtain four independent equations:
\begin{eqnarray}\label{pe12}
 &&N\left(\partial_{\eta}^{2} + k^{2} + i\frac{a^{'}}{a}\gamma_{0}\gamma_{i}k^{i} + m^{2}a^{2} - \frac{3}{4}\frac{a^{'2}}{a^{2}}\right)g^{(+)}(\eta)\lambda^{S}_{\{\pm,\mp\}}({\bf k})e^{i{\bf k \cdot x}} = 0\nonumber\\
 &&N\left(\partial_{\eta}^{2} + k^{2} - i\frac{a^{'}}{a}\gamma_{0}\gamma_{i}k^{i} + m^{2}a^{2} - \frac{3}{4}\frac{a^{'2}}{a^{2}}\right)g^{(-)}(\eta)\lambda^{A}_{\{\pm,\mp\}}({\bf k})e^{-i{\bf k \cdot x}} = 0.
\end{eqnarray}
In the Weyl representation for the Dirac gamma matrices, the term $\gamma_{0}\gamma_{i}k^{i}$ is
\begin{equation}\nonumber
 \gamma_{0}\gamma_{i}k^{i} = \left( \begin{array}{cccc}        -k_{3} & -k_{1}+ik_{2} & 0            & 0 \\
                                                        -k_{1}-ik_{2} & k_{3}        & 0            & 0 \\
                                                                    0 & 0            & k_{3}        & k_{1}-ik_{2} \\
                                                                    0 & 0            & k_{1} +ik_{2}& -k_{3} 
                                      \end{array} \right).
\end{equation}
Thus, the above system can be written in more compact form as
\begin{eqnarray}\label{pe13}
 && ND^ {(+)}g^{(+)}(\eta)\lambda_{\{\pm,\mp\}}^{S}({\bf k})e^{i{\bf k \cdot x}} = 0,\nonumber\\
 && ND^ {(-)}g^{(-)}(\eta)\lambda_{\{\pm,\mp\}}^{A}({\bf k})e^{-i{\bf k \cdot x}} = 0,
\end{eqnarray}
where $D^{(\pm)}$ are the matrices
\begin{equation}\nonumber
D^{(\pm)} =  \left( \begin{array}{cccc} D^{(\pm)}_{00} & D^{(\pm)}_{01} & 0            & 0 \\
                                                       D^{(+)}_{10} & D^{(\pm)}_{11} & 0             &  0 \\
                                                 0 & 0            & D^{(\pm)}_{22} &  D^{(\pm)}_{23}\\
                                                                 0 & 0            &   D^{(\pm)}_{32}    & D^{(\pm)}_{33}
                                      \end{array} \right)
\end{equation}
with components:
\[
D^{(+)}_{00}=D^{(+)}_{33}=D^{(-)}_{11}=D^{(-)}_{22}=\partial_{\eta}^{2} + k^{2} + m^{2}a^{2} - \frac{3}{4}\frac{a^{'2}}{a^{2}} -i\frac{a^{'}}{a}k_{3}
\]
\[
D^{(+)}_{11}=D^{(+)}_{22}=D^{(-)}_{00}=D^{(-)}_{33}=\partial_{\eta}^{2} + k^{2} + m^{2}a^{2} - \frac{3}{4}\frac{a^{'2}}{a^{2}} +i\frac{a^{'}}{a}k_{3}
\]
\[
D^{(+)}_{01}=-D^{(+)}_{23}= -D^{(-)}_{01}=D^{(-)}_{23}=- i\frac{a^{'}}{a}k_{1}-\frac{a^{'}}{a}k_{2}
\]
\[
D^{(+)}_{10}=-D^{(+)}_{32}=-D^{(-)}_{10}=D^{(-)}_{32}=-i\frac{a^{'}}{a}k_{1} + \frac{a^{'}}{a}k_{2}
\]
By assuming that ${\lambda^{S}_{\{\pm,\mp\}}}$ and ${\lambda^{A}_{\{\pm,\mp\}}}$ can be written in the general form (see Appendix A):
\begin{equation}
 \lambda^{S}_{\{ \pm,\mp \}}({\bf k})= \left( \begin{array}{c}  b^{1,2} \\ c^{1,2} \\ d^{1,2} \\ f^{1,2} \end{array} \right)\hspace{1cm} \lambda^{A}_{\{ \pm,\mp \}}({\bf k})= \left( \begin{array}{c}  h^{1,2} \\ l^{1,2} \\ m^{1,2} \\ n^{1,2} \end{array} \right)
\end{equation}
with all the components non null, where the superscript $(1,2)$ stands for the helicities $\{+,-\}$ and $\{-,+\}$, respectively, the system (\ref{pe13}) can be put into the following form:
\begin{eqnarray}\nonumber
&&\left(\partial_{\eta}^{2} + k^{2} + m^{2}a^{2} - \frac{3}{4}\frac{a^{'2}}{a^{2}} + \frac{a^{'}}{a}\left[(-ik_{1}-k_{2})\frac{c^{1,2}}{b^{1,2}}- ik_{3}\right]\right)g^{(+)}(\eta) = 0,\\\nonumber 
&&\left(\partial_{\eta}^{2} + k^{2} + m^{2}a^{2} - \frac{3}{4}\frac{a^{'2}}{a^{2}} + \frac{a^{'}}{a}\left[(-ik_{1}+k_{2})\frac{b^{1,2}}{c^{1,2}}+ ik_{3}\right]\right)g^{(+)}(\eta) = 0,\\\nonumber 
&&\left(\partial_{\eta}^{2} + k^{2} + m^{2}a^{2} - \frac{3}{4}\frac{a^{'2}}{a^{2}} + \frac{a^{'}}{a}\left[(+ik_{1}+k_{2})\frac{f^{1,2}}{d^{1,2}}+ ik_{3}\right]\right)g^{(+)}(\eta) = 0,\\\nonumber 
&&\left(\partial_{\eta}^{2} + k^{2} + m^{2}a^{2} - \frac{3}{4}\frac{a^{'2}}{a^{2}} + \frac{a^{'}}{a}\left[(+ik_{1}-k_{2})\frac{d^{1,2}}{f^{1,2}}- ik_{3}\right]\right)g^{(+)}(\eta) = 0,\\\nonumber 
&&\left(\partial_{\eta}^{2} + k^{2} + m^{2}a^{2} - \frac{3}{4}\frac{a^{'2}}{a^{2}} + \frac{a^{'}}{a}\left[(+ik_{1}+k_{2})\frac{l^{1,2}}{h^{1,2}}+ ik_{3}\right]\right)g^{(-)}(\eta) = 0,\\\nonumber 
&&\left(\partial_{\eta}^{2} + k^{2} + m^{2}a^{2} - \frac{3}{4}\frac{a^{'2}}{a^{2}} + \frac{a^{'}}{a}\left[(+ik_{1}-k_{2})\frac{h^{1,2}}{l^{1,2}}- ik_{3}\right]\right)g^{(-)}(\eta) = 0,\\\nonumber 
&&\left(\partial_{\eta}^{2} + k^{2} + m^{2}a^{2} - \frac{3}{4}\frac{a^{'2}}{a^{2}} + \frac{a^{'}}{a}\left[(-ik_{1}-k_{2})\frac{n^{1,2}}{m^{1,2}}- ik_{3}\right]\right)g^{(-)}(\eta) = 0,\\\nonumber 
&&\left(\partial_{\eta}^{2} + k^{2} + m^{2}a^{2} - \frac{3}{4}\frac{a^{'2}}{a^{2}} + \frac{a^{'}}{a}\left[(-ik_{1}+k_{2})\frac{m^{1,2}}{n^{1,2}}+ ik_{3}\right]\right)g^{(-)}(\eta) = 0,\nonumber 
\end{eqnarray}
which can be reduced to a single equation
\begin{equation}\label{pe14}
\left(\partial_{\eta}^{2} + k^{2} + m^{2}a^{2} - \frac{3}{4}\frac{a^{'2}}{a^{2}} + \frac{a^{'}}{a}[X_{j} + iY_{j}]\right)g(\eta) = 0,
\end{equation}
where ${X_{j} = X_{j}(k_{1},k_{2}, k_{3})}$ and ${Y_{j} = Y_{j}(k_{1},k_{2},k_{3})}$ with ${j=}$ 1, 2, 3, 4 for ${g^{(+)}(\eta)}$ of ${\Lambda_{\{+,-\}}^{S}}$; ${j=}$ 5, 6, 7, 8 for ${g^{(+)}(\eta)}$ of ${\Lambda_{\{-,+\}}^{S}}$;
${j=}$ 9, 10, 11, 12 for ${g^{(-)}(\eta)}$ of ${\Lambda_{\{+,-\}}^{A}}$ and ${j=}$ 13, 14, 15, 16 for ${g^{(-)}(\eta)}$ of ${\Lambda_{\{-,+\}}^{A}}$. 

Equation (\ref{pe14}) is the main equation to start the study of particle creation for the Elko field. Several interesting consequences can be obtained from (\ref{pe14}). First, if we use spherical coordinates into the term $[X_{j} + iY_{j}]$, it can be showed that $X_j=0$ for all components and $Y_j=\mp k =\mp \vert {\bf k}\vert$, being $-k$ for ${\Lambda^{S}_{\{+,-\}}}$, ${\Lambda^{A}_{\{-,+\}}}$ and ${+k}$ for ${\Lambda^{S}_{\{-,+\}}}$, ${\Lambda^{A}_{\{+,-\}}}$. The last two terms into curl brackets, namely the terms proportional to $a'^2$ and $a'$, are specific to the Elko field. The first one carries the spinor coupling to the gravitational field that came from the spin connection $\Gamma_\mu$ and the second carries the explicit spinorial structure of the field. When both terms are discard we obtain exactly the same equation for the standard scalar field \cite{davies}. 

In which follows we will keep the general components $[X_{j} + iY_{j}]$ in arbitrary coordinate system, and at the end we will take the spherical coordinate system for a explicit calculation.

\section{Solutions for an asymptotically flat metric}

Now we apply the above discussion  to a particular metric that is asymptotically flat in the past and in the future: 
\begin{equation}\label{pe15}
a^{2}(\eta) = A + B\tanh(\rho\eta), 
\end{equation}
where ${A}$, ${B}$ e ${\rho}$ are real, with ${A > B > 0}$ and ${\rho > 0}$. Such metric admits analytic solutions for both scalar field and Dirac fermionic fields, thus it is possible to compare the final results. In the limits ${\eta\longrightarrow\pm\infty}$ such metric satisfies: 
\begin{eqnarray}
&& \frac{a^{'}}{a} = \frac{a^{'2}}{a^{2}} \longrightarrow 0,\nonumber\\
&& a^{2}(\eta) \longrightarrow A \pm B.\label{asymp}
\end{eqnarray}
Defining an effective mass as
\begin{eqnarray}
&& m_{eff}^{2}(\eta) = m^{2}a^{2} - \frac{3}{4}\frac{a^{'2}}{a^{2}} + \frac{a^{'}}{a}(X_{j} + iY_{j}),\label{pe16}
\end{eqnarray}
we have the frequency $\omega^{2}(\eta) = k^{2} + m_{eff}^{2}(\eta)$, so that Eq. (\ref{pe14}) simplifies to:
\begin{equation}\label{pe14s}
\left(\partial_{\eta}^{2} +\omega^{2}(\eta) \right)g(\eta) = 0,
\end{equation}
in analogy to a frequency time varying harmonic oscillator. The frequencies in the past and in the future can be defined  as
\begin{equation}
\omega(\eta)  \equiv \Bigg\{\begin{array}{cc} \omega_{in} = \sqrt{k^{2} + m^{2}(A - B)}, \qquad\eta\longrightarrow -\infty,\\
 \omega_{out} = \sqrt{k^{2} + m^{2}(A + B)}, \qquad\eta\longrightarrow +\infty.
                     \end{array} \label{pe17}
\end{equation}

Substituting (\ref{pe15}) into (\ref{pe14}) we have:
\begin{eqnarray}
&& \Bigg[\partial_{\eta}^{2} + k^{2} + m^{2}(A + B\tanh(\rho\eta)) - \frac{3}{4}\cdot\frac{1}{4}(B\rho)^{2}\left(\frac{1-\tanh^{2}(\rho\eta)}{A+B\tanh(\rho\eta)}\right)^{2}\nonumber\\\label{pe18}
&&+ \frac{1}{2}B\rho\left(\frac{1-\tanh^{2}(\rho\eta)}{A+B\tanh(\rho\eta)}\right)(X_{j} + iY_{j})\Bigg]g(\eta) = 0\label{pe18}
\end{eqnarray}

\begin{figure*}[tb]
\begin{center}
\centering
\includegraphics[width=0.4\textwidth]{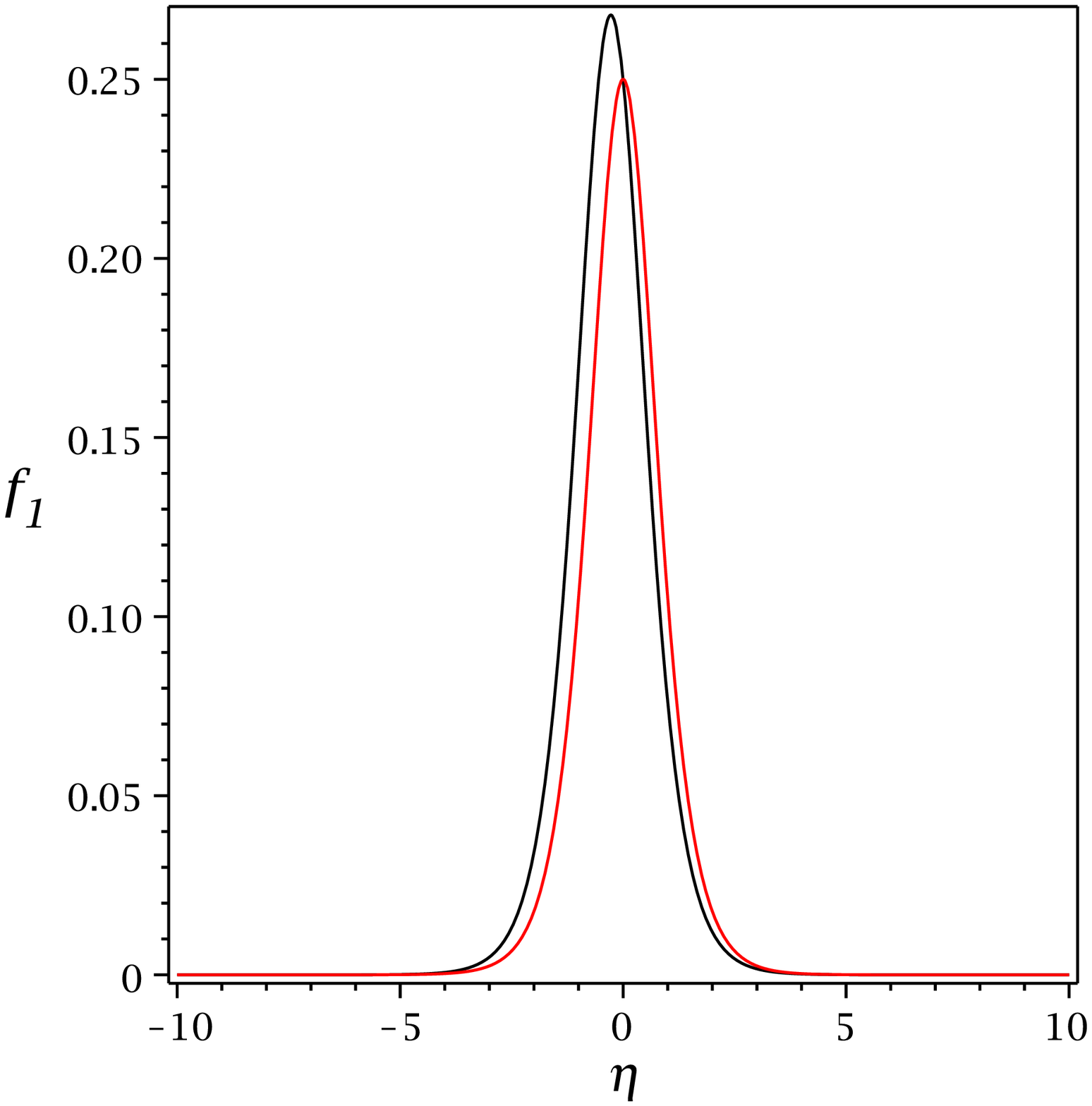}
\hspace{1.3cm}
\includegraphics[width=0.4\textwidth]{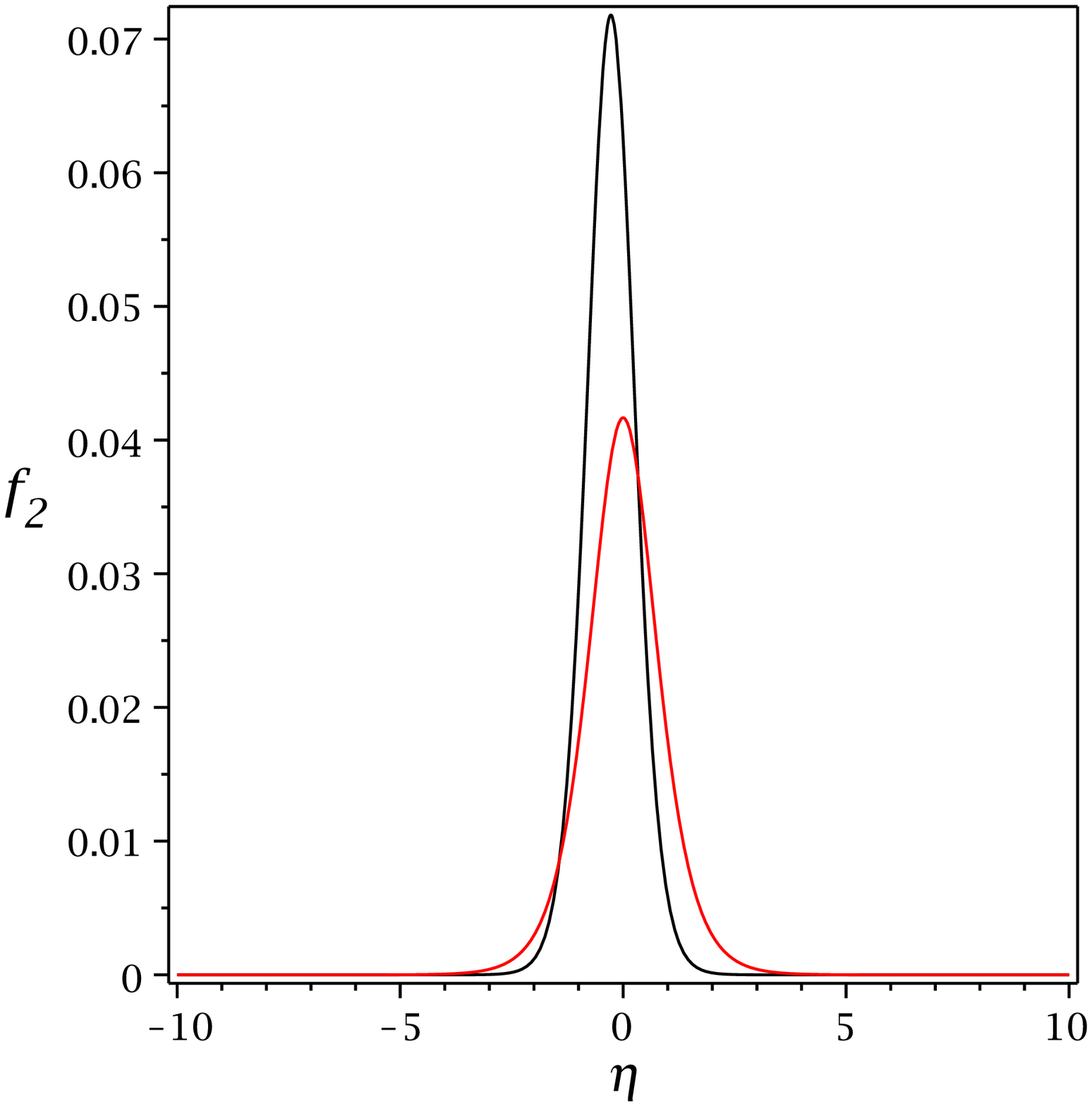}\\
\hspace{1.2cm}(a)\hspace{7.7cm}(b)\\
\vspace{0.3cm}
\includegraphics[width=0.4\textwidth]{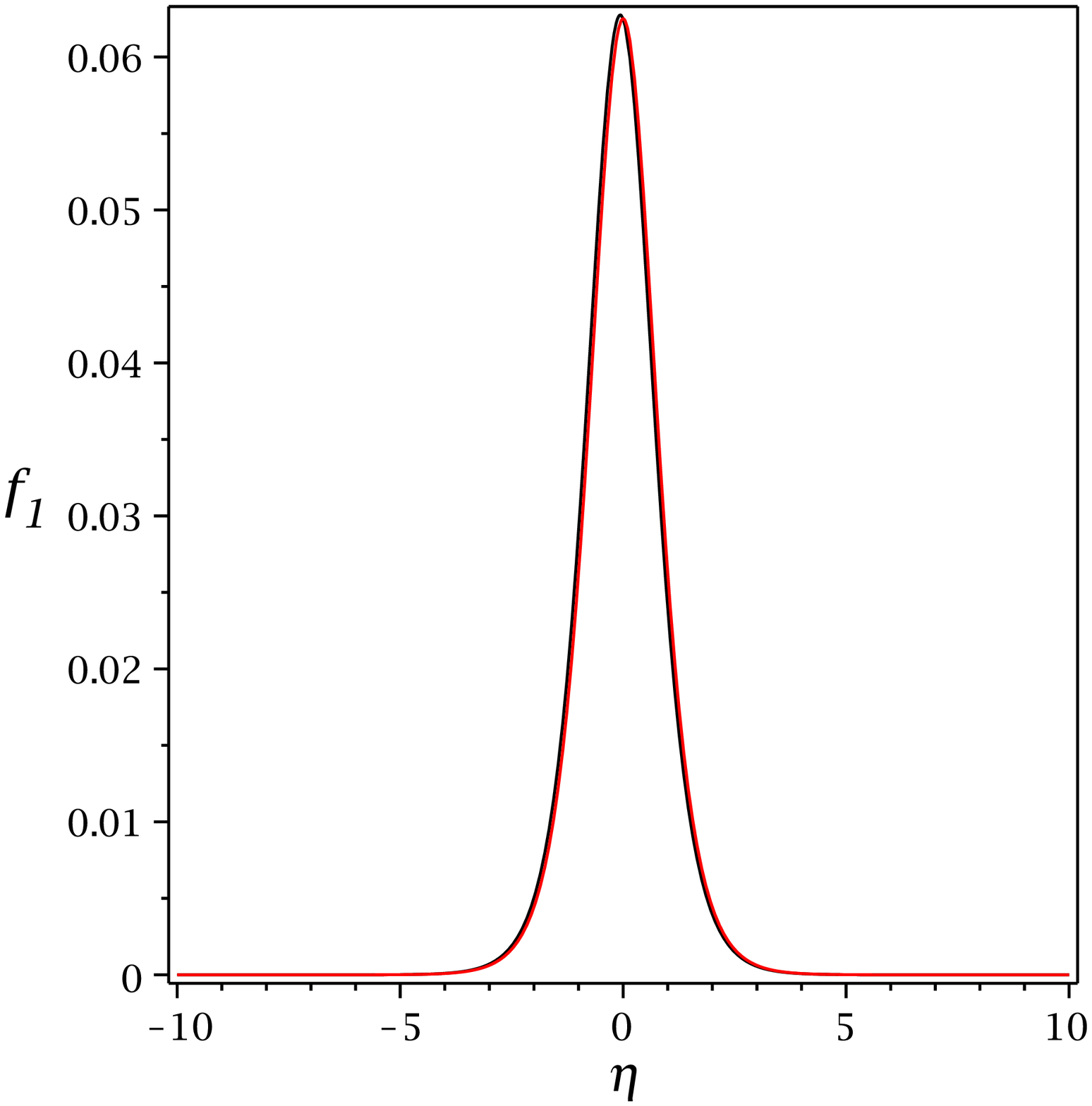}
\hspace{1.3cm}
\includegraphics[width=0.4\textwidth]{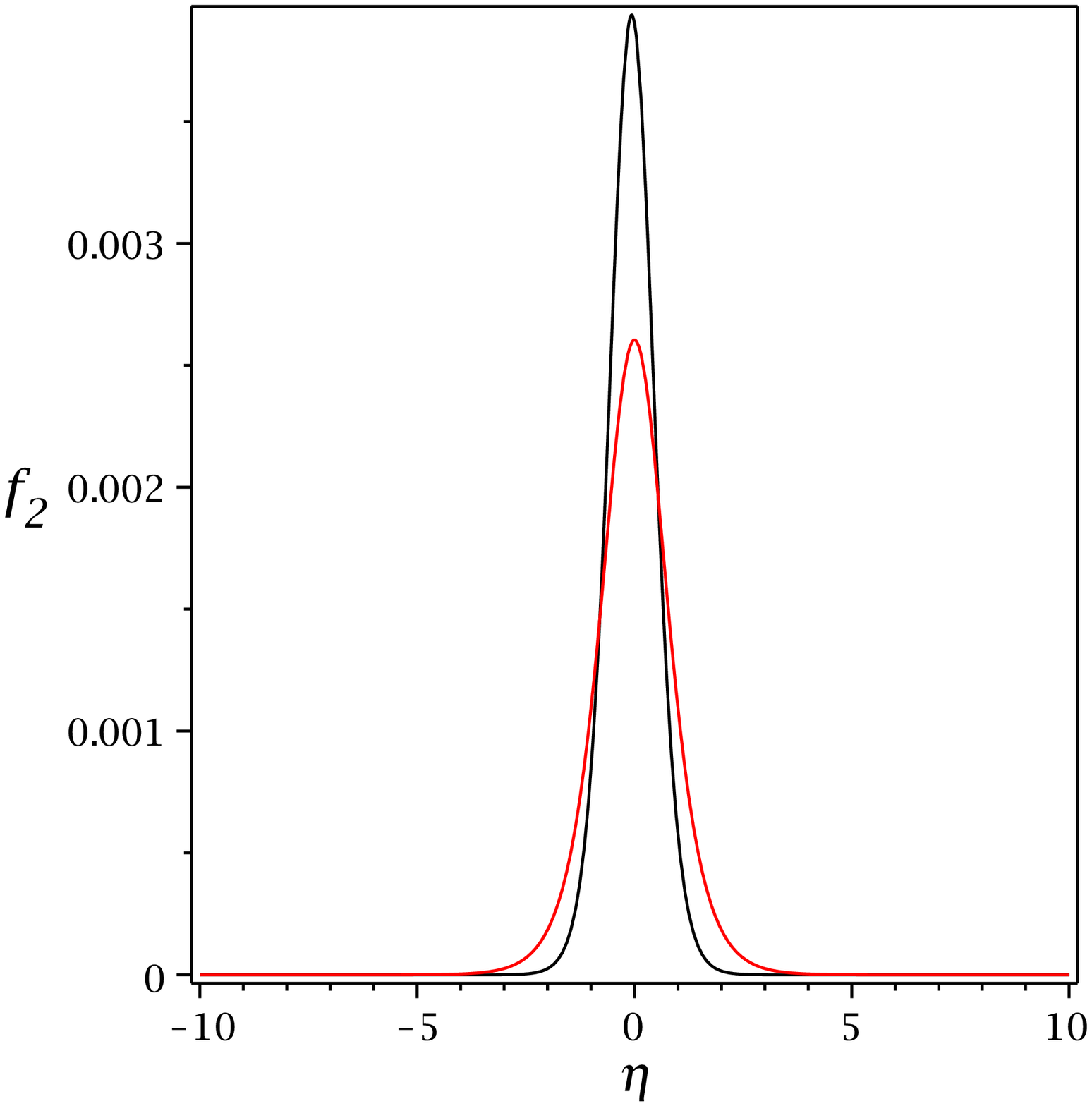}\\
\hspace{1.2cm}(c)\hspace{7.7cm}(d)
\end{center}
\caption{Graphical analysis for the exact functions ($f_{1\,ex},\;f_{2\,ex}$), in black lines and approximated functions ($f_{1\,ap},\;f_{2\,ap}$), in red lines, from Eqs. (\ref{pe21}) and (\ref{pe22}) for $c_1=(A+B)/A$ and $c_2=(2/3)c_1^2$, with the following values of parameters: (a) $A=2$, $B=1$;  (b) $A=2$, $B=1$; (c) $A=8$, $B=1$;  (d) $A=8$, $B=1$.}\label{fig1}
\end{figure*}

It is important to stop for a moment to make some considerations about the differential equation (\ref{pe18}). The last two terms within the square brackets carries the explicit contribution of the Elko spinor, coming from $a'^2$ and $a'$ as previously mentioned. However this differential equation has no known analytical solutions. Although the asymptotic behaviour (\ref{asymp}) of the metric allows to discard these terms in the limits ${\eta\longrightarrow\pm\infty}$, it is well  known that the process of creation in Parker's formalism compares the number of created particles  between the asymptotic stages, thus we would like to take into account the explicit contribution of these terms, otherwise we would simply have the same rate of creation from the standard scalar field. This can be done by substituting those terms by some function with the same behaviour in the range. A function that preserves the same shape of the previous one will contribute with nearly the same quantity into the final calculations, since that the total of created particles is just taken in the limit of asymptotic time from $-\infty$ to $+\infty$. We use the following approximations:
\begin{eqnarray}\label{pe21}
\frac{a^{'}}{a} &=&f_{1\,ex}= \frac{1}{2}\frac{B\rho(1-\tanh^{2}(\rho\eta))}{(A+B\tanh(\rho\eta))} \approx \frac{1}{2}c_{1}\frac{B\rho(1-\tanh^{2}(\rho\eta))}{(A+B)}                =f_{1\,ap}
\end{eqnarray}
\begin{eqnarray}\label{pe22}
\frac{a^{'2}}{a^{2}} &=& f_{2\,ex}=\frac{1}{4}\frac{B^{2}\rho^{2}(1-\tanh^{2}(\rho\eta))^{2}}{(A+B\tanh(\rho\eta))^{2}} \approx \frac{1}{4}\frac{c_{2}B^{2}\rho^{2}(1-\tanh^{2}(\rho\eta))}{(A+B)^{2}}=f_{2\,ap}
\end{eqnarray}
where $c_1$ and $c_2$ are constants to be adjusted in the approximated functions, in order to maintain the area below the function to the same order. A graphical and numerical analysis of the exact functions ($f_{1\,ex},\;f_{2\,ex}$) and approximated functions ($f_{1\,ap},\;f_{2\,ap}$) suggest a reasonable choice for the parameters as $c_1=(A+B)/A$ and $c_2=(2/3)c_1^2$. Some examples are plotted in the Figure \ref{fig1} for different values of the constants $A$ and $B$. We have verified that the functions $f_{1\,ex}$ and $f_{1\,ap}$ are nearly equal to all values of $A$ and $B$ satisfying $A/B>2$. The difference between the area below the curves are less than $9\%$ for $A/B>2$ and less than $1\%$ for $A/B>6$. For the functions $f_{2\,ex}$ and $f_{2\,ap}$ the difference in the amplitude are more evident, however we have verified numerically that the difference between the area below the curves are less than $7\%$ for $A/B>3$ and less than $1\%$ for $A/B>8$. Thus we conclude that our approximation is good enough to ensure an error less than $1\%$ for both functions if the condition $A/B>8$ be satisfied. 

Substituting the approximated functions into (\ref{pe18}) and making the change of variables ${\xi = [1+\tanh(\rho\eta)]/{2}}$ we obtain, after a division by ${4\rho^{2}\xi^{2}(1-\xi)^{2}}$,
\begin{eqnarray}
&&\Bigg[\frac{d^{2}}{d\xi^{2}} + \left(\frac{1}{\xi-1} + \frac{1}{\xi}\right)\frac{d}{d\xi} + \left(\frac{\omega_{out}^{2}}{4\rho^{2}}\frac{1}{\xi-1} - \frac{\omega_{in}^{2}}{4\rho^{2}}\frac{1}{\xi}\right)\cdot\frac{1}{\xi(\xi-1)}\nonumber\\
&&+\frac{B}{8(A+B)}\left(\frac{3c_{2}B}{2(A+B)} - \frac{4c_{1}(X_{j} + iY_{j})}{\rho}\right)\cdot\frac{1}{\xi(\xi-1)}\Bigg]\tilde{g}(\xi) = 0.\label{pe23}
\end{eqnarray}
Such second order differential equation is of Riemann-Papperitz type, whose solutions are generalizations of hypergeometric functions. This shows the advantage to use the approximated functions (\ref{pe21}) and (\ref{pe22}), since that now we have analytic solutions and the informations about the Elko field will be present into the solutions.

The solution for ${\tilde{g}(\xi)}$ from (\ref{pe23}) is (see Appendix B):
\begin{equation}\label{pe24}
\tilde{g}(\xi) = \xi^{\frac{i\omega_{in}}{2\rho}}(1-\xi)^{\frac{i\omega_{out}}{2\rho}}P\begin{Bmatrix} 
0 & 1 & \infty \\ 
0 & 0 & \frac{1\pm\sqrt{1-4r}}{2} + \frac{i\omega_{+}}{\rho} & \xi\\ 
-\frac{i\omega_{in}}{\rho} & -\frac{i\omega_{out}}{\rho} & \frac{1\mp\sqrt{1-4r}}{2} + \frac{i\omega_{+}}{\rho}  
\end{Bmatrix}
\end{equation}
where ${r = \frac{B}{8(A+B)}\left(\frac{3c_{2}B}{2(A+B)} - \frac{4c_{1}(X_{j} + iY_{j})}{\rho}\right)}$ and $P\{ \}$ is called Riemann symbol. Notice that, from the discussion that follows (\ref{pe14}), in spherical coordinates we have $Y_j=\mp k =$, being $-k$ for ${\Lambda^{S}_{\{+,-\}}}$, ${\Lambda^{A}_{\{-,+\}}}$ and ${+k}$ for ${\Lambda^{S}_{\{-,+\}}}$, ${\Lambda^{A}_{\{+,-\}}}$. Thus we will refer to a single $r$ coefficient but remembering that it can assume two distinct values, namely $r=r_\mp$, according to signal of $Y_j$.

The solution (\ref{pe24}) can be expressed in terms of hypergeometric functions $_2F_1(a,b;c;z)$. We let to Appendix B the cumbersome manipulations involving properties of hypergeometric functions, starting from the Riemann-Papperitz differential equation.

In order to study the particle creation between the asymptotic limits $-\infty < \eta < \infty$, here named \emph{in} and \emph{out} states, we need the functions $g^{(\pm)}(\eta)$ for these states. From (\ref{pe37}) we obtain:
\begin{eqnarray}\label{pe38}
g_{in}^{(+)}(\eta) &=& \exp\left[i\left(-\omega_{+}\eta - \frac{\omega_{-}}{\rho}\ln(2\cosh(\rho\eta))\right)\right]\\\nonumber
                                         &\times&_{2}F_{1}\left(\frac{1\mp(\sqrt{1-4r})^{*}}{2} + \frac{i\omega_{-}}{\rho}, \frac{1\pm(\sqrt{1-4r})^{*}}{2} + \frac{i\omega_{-}}{\rho};1-\frac{i\omega_{in}}{\rho};\frac{1+\tanh(\rho\eta)}{2}\right),\nonumber 
\end{eqnarray}
\begin{eqnarray}\label{pe39}
g_{in}^{(-)}(\eta) &=& \exp\left[i\left(-\omega_{-}\eta - \frac{\omega_{+}}{\rho}\ln(2\cosh(\rho\eta))\right)\right]\\\nonumber
                                         &\times&_{2}F_{1}\left(\frac{1\mp(\sqrt{1-4r})^{*}}{2} + \frac{i\omega_{+}}{\rho}, \frac{1\pm(\sqrt{1-4r})^{*}}{2} + \frac{i\omega_{+}}{\rho};1+\frac{i\omega_{in}}{\rho};\frac{1+\tanh(\rho\eta)}{2}\right)\nonumber 
\end{eqnarray}
from which follows that: ${g^{(\pm)}_{in}(\eta\to -\infty) = e^{\mp i\omega_{in}\eta}}$ and:
\begin{eqnarray}\label{pe53}
g_{out}^{(+)}(\eta) &=& \exp\left[i\left(-\omega_{+}\eta - \frac{\omega_{-}}{\rho}\ln(2\cosh(\rho\eta))\right)\right]\\\nonumber
            &\times&_{2}F_{1}\left(\frac{1\mp(\sqrt{1-4r})^{*}}{2} + \frac{i\omega_{-}}{\rho}, \frac{1\pm(\sqrt{1-4r})^{*}}{2} + \frac{i\omega_{-}}{\rho};\frac{i\omega_{out}}{\rho};\frac{1-\tanh(\rho\eta)}{2}\right),\nonumber
\end{eqnarray}
\begin{eqnarray}\label{pe54}
g_{out}^{(-)}(\eta) &=& \exp\left[i\left(\omega_{-}\eta + \frac{\omega_{+}}{\rho}\ln(2\cosh(\rho\eta))\right)\right]\\\nonumber
            &\times&_{2}F_{1}\left(\frac{1\mp(\sqrt{1-4r})^{*}}{2} - \frac{i\omega_{+}}{\rho}, \frac{1\pm(\sqrt{1-4r})^{*}}{2} - \frac{i\omega_{+}}{\rho};-\frac{i\omega_{out}}{\rho};\frac{1-\tanh(\rho\eta)}{2}\right),\nonumber
\end{eqnarray}
with ${g^{(\pm)}_{out}(\eta\to \infty) = \e^{\mp i\omega_{out}\eta}}$. Thus we can write the $g^{\pm}_{in}(\eta\to -\infty)$ states in terms of  $g^{\pm}_{out}(\eta\to \infty)$,
\begin{equation}\label{pe51}
g_{in}^{(+)}(\eta) = \sigma \e^{-i\omega_{out}\eta} + \varrho \e^{i\omega_{out}\eta},
\end{equation}
\begin{equation}\label{pe52}
g_{in}^{(-)}(\eta) = \pi \e^{-i\omega_{out}\eta} + \tau \e^{i\omega_{out}\eta}, 
\end{equation}
where ${\sigma}$, ${\varrho}$, ${\pi}$ and ${\tau}$ are the Bogolyubov coefficients:
\begin{equation}\label{pe55}
\sigma = C_{1} = \frac{\Gamma\left(1-\frac{i\omega_{in}}{\rho}\right)\Gamma\left(-\frac{i\omega_{out}}{\rho}\right)}{\Gamma\left(\frac{1\pm(\sqrt{1-4r})^{*}}{2} - \frac{i\omega_{+}}{\rho}\right)\Gamma\left(\frac{1\mp(\sqrt{1-4r})^{*}}{2} - \frac{i\omega_{+}}{\rho}\right)},
\end{equation}\\
\begin{equation}\label{pe56}
\varrho = C_{2} = \frac{\Gamma\left(1-\frac{i\omega_{in}}{\rho}\right)\Gamma\left(\frac{i\omega_{out}}{\rho}\right)}{\Gamma\left(\frac{1\mp(\sqrt{1-4r})^{*}}{2} + \frac{i\omega_{-}}{\rho}\right)\Gamma\left(\frac{1\pm(\sqrt{1-4r})^{*}}{2} + \frac{i\omega_{-}}{\rho}\right)},
\end{equation}\\
\begin{equation}\label{pe57}
\pi = C_{3} = \frac{\Gamma\left(1+\frac{i\omega_{in}}{\rho}\right)\Gamma\left(-\frac{i\omega_{out}}{\rho}\right)}{\Gamma\left(\frac{1\pm(\sqrt{1-4r})^{*}}{2} - \frac{i\omega_{-}}{\rho}\right)\Gamma\left(\frac{1\mp(\sqrt{1-4r})^{*}}{2} - \frac{i\omega_{-}}{\rho}\right)}, 
\end{equation}\\
\begin{equation}\label{pe58}
\tau = C_{4} = \frac{\Gamma\left(1+\frac{i\omega_{in}}{\rho}\right)\Gamma\left(\frac{i\omega_{out}}{\rho}\right)}{\Gamma\left(\frac{1\mp(\sqrt{1-4r})^{*}}{2} + \frac{i\omega_{+}}{\rho}\right)\Gamma\left(\frac{1\pm(\sqrt{1-4r})^{*}}{2} + \frac{i\omega_{+}}{\rho}\right)}. 
\end{equation}

Finally, the four Elko fields $\Lambda = \{ \Lambda_{\beta}^{S}, \Lambda_{\beta}^{A}\}$ from (\ref{pe11}) are completely determined for \emph{in} and \emph{out} regions.

\section{Quantum field and creation of Elko particles}

Now we proceed to the quantization of the field and the explicit calculation of the number of created particles.

The two complete basis of solutions for self-conjugate and anti-self-conjugate Elko field in the regions
{\it in} and {\it out} are, respectively, ${\{\Lambda^{S}_{in}({\bf k}, \beta,\eta),\Lambda^{A}_{in}({\bf k}, \beta,\eta)\}}$ e
${\{\Lambda^{S}_{out}({\bf k}, \beta,\eta),\Lambda^{A}_{out}({\bf k}, \beta,\eta)\}}$, with ${\beta}$ denoting the helicities. The Fourier decomposition of the field is (notice from (\ref{pe11}) the presence of a normalization factor $N$):
\begin{equation}\label{pe61}
\mathfrak{f}_{in}(x) = \int d^{3}k\sum\limits_{\beta}[\hat{a}_{in}({\bf k},\beta)\Lambda_{in}^{S}({\bf k},\beta,\eta)e^{i{\bf k\cdot x}} + \hat{b}_{in}^{\dag}({\bf k},\beta)\Lambda_{in}^{A}({\bf k},\beta,\eta)e^{-i{\bf k\cdot x}}],
\end{equation}
\begin{equation}\label{pe62}
\mathfrak{f}_{out}(x) = \int d^{3}k\sum\limits_{\beta}[\hat{a}_{out}({\bf k},\beta)\Lambda_{out}^{S}({\bf k},\beta,\eta)e^{i{\bf k\cdot x}} + \hat{b}_{out}^{\dag}({\bf k},\beta)\Lambda_{out}^{A}({\bf k},\beta,\eta)e^{-i{\bf k\cdot x}}],
\end{equation}
with the creation ($\hat{a}_{in/out}^{\dag},\hat{b}_{in/out}^{\dag}$) and annihilation ($\hat{a}_{in/out}, \hat{b}_{in/out}$) operators satisfying:
\begin{eqnarray}\label{pe63}
&&\{\hat{a}_{in/out}({\bf k},\beta),\hat{a}_{in/out}^{\dag}({\bf k^{'}},\beta^{'})\} = \delta^{3}({\bf k} - {\bf k^{'}})\delta_{\beta\beta^{'}},\nonumber\\
&&\{\hat{b}_{in/out}({\bf k},\beta),\hat{b}_{in/out}^{\dag}({\bf k^{'}},\beta^{'})\} = \delta^{3}({\bf k} - {\bf k^{'}})\delta_{\beta\beta^{'}},\nonumber\\
&&\{\hat{a}_{in/out}({\bf k},\beta),\hat{a}_{in/out}({\bf k^{'}},\beta^{'})\} = \{\hat{a}_{in/out}^{\dag}({\bf k},\beta),\hat{a}_{in/out}^{\dag}({\bf k^{'}},\beta^{'})\} = 0,\nonumber\\
&&\{\hat{b}_{in/out}({\bf k},\beta),\hat{b}_{in/out}({\bf k^{'}},\beta^{'})\} = \{\hat{b}_{in/out}^{\dag}({\bf k},\beta),\hat{b}_{in/out}^{\dag}({\bf k^{'}},\beta^{'})\} = 0.
\end{eqnarray}

By means of Bogolyubov transformation it is possible to describe the modes of solutions {\it in} in terms of {\it out}, analogous to ${g_{in}^{(\pm)}(\eta)}$ in terms of ${g_{out}^{(\pm)}(\eta)}$:
\begin{eqnarray}\label{pe64}
&& \Lambda_{in}^{S}({\bf k}, \beta,\eta) = \sum\limits_{\beta^{'}}[\sigma_{\beta\beta^{'}}\Lambda_{out}^{S}({\bf k},\beta^{'},\eta) + \varrho_{\beta\beta^{'}}\Lambda_{out}^{A}({\bf k},\beta^{'},\eta)],\nonumber\\
&& \Lambda_{in}^{A}({\bf k}, \beta,\eta) = \sum\limits_{\beta^{'}}[\pi_{\beta\beta^{'}}\Lambda_{out}^{S}({\bf k},\beta^{'},\eta) + \tau_{\beta\beta^{'}}\Lambda_{out}^{A}({\bf k},\beta^{'},\eta)],
\end{eqnarray}
where ${\sigma_{\beta\beta^{'}}}$, ${\varrho_{\beta\beta^{'}}}$, ${\pi_{\beta\beta^{'}}}$ e ${\tau_{\beta\beta^{'}}}$ are the Bogolyubov coefficients. The operators can be written as:
\begin{eqnarray}\label{pe65}
&&\hat{a}_{out}({\bf k},\beta) = \sum\limits_{\beta^{'}}[\sigma_{\beta\beta^{'}}\hat{a}_{in}({\bf k},\beta^{'}) + \pi_{\beta\beta^{'}}\hat{b}^{\dag}_{in}({\bf k},\beta^{'})],\\\nonumber
&&\hat{b}_{out}({\bf k},\beta) = \sum\limits_{\beta^{'}}[\tau^{*}_{\beta\beta^{'}}\hat{b}_{in}({\bf k},\beta^{'}) + \varrho^{*}_{\beta\beta^{'}}\hat{a}^{\dag}_{in}({\bf k},\beta^{'})].\nonumber
\end{eqnarray}
In order to satisfy (\ref{pe63}) we must have:
\begin{eqnarray}\label{pe67}
\sum\limits_{\beta^{'}}(\lvert \sigma_{\beta\beta^{'}}\lvert^{2} + \lvert \pi_{\beta\beta^{'}}\lvert^{2}) = 1,\nonumber\\
\sum\limits_{\beta^{'}}(\lvert \tau_{\beta\beta^{'}}\lvert^{2} + \lvert \varrho_{\beta\beta^{'}}\lvert^{2}) = 1.
\end{eqnarray}

The expected value of created self-conjugated Elko particles for an observer at {\it out} in terms of the {\it in} vacuum is:
\begin{eqnarray}\label{pe68}
\langle N^{S}\rangle &=& \langle0_{in}|\hat{a}^{\dag}_{out}({\bf k},\beta)\hat{a}_{out}({\bf k},\beta)|0_{in}\rangle\nonumber\\
&=&\langle0_{in}|\sum\limits_{\beta^{'}}(\sigma^{*}_{\beta\beta^{'}}\hat{a}^{\dag}_{in}({\bf k},\beta^{'}) + \pi^{*}_{\beta\beta^{'}}\hat{b}_{in}({\bf k},\beta^{'}))\times(\sigma_{\beta\beta^{'}}\hat{a}_{in}({\bf k},\beta^{'}) + \pi_{\beta\beta^{'}}\hat{b}^{\dag}_{in}({\bf k},\beta^{'}))|0_{in}\rangle\nonumber\\
&=&\sum\limits_{\beta^{'}}\lvert\pi_{\beta\beta^{'}}\rvert^{2}.
\end{eqnarray}
At the same time, the expected value of created anti-self-conjugated Elko particles for an observer at {\it out} in terms of the {\it in} vacuum is:
\begin{eqnarray}\label{pe69}
\langle N^{A}\rangle &=& \langle0_{in}|\hat{b}^{\dag}_{out}({\bf k},\beta)\hat{b}_{out}({\bf k},\beta)|0_{in}\rangle\nonumber\\
&=&\langle0_{in}|\sum\limits_{\beta^{'}}(\tau_{\beta\beta^{'}}\hat{b}^{\dag}_{in}({\bf k},\beta^{'}) + \varrho_{\beta\beta^{'}}\hat{a}_{in}({\bf k},\beta^{'}))\times(\tau^{*}_{\beta\beta^{'}}\hat{b}_{in}({\bf k},\beta^{'}) + \varrho^{*}_{\beta\beta^{'}}\hat{a}^{\dag}_{in}({\bf k},\beta^{'}))|0_{in}\rangle\nonumber\\
&=&\sum\limits_{\beta^{'}}\lvert\varrho_{\beta\beta^{'}}\rvert^{2}.
\end{eqnarray}
Thus, the total expected value of created Elko particles for an observer at {\it out} in terms of the {\it in} vacuum is:
\begin{equation}\label{pe70}
\langle N_{\lambda}\rangle = \langle N^{S}\rangle + \langle N^{A}\rangle = \sum\limits_{\beta^{'}}(\lvert\pi_{\beta\beta^{'}}\rvert^{2}+\lvert\varrho_{\beta\beta^{'}}\rvert^{2}). 
\end{equation}
\indent The modes ${\{\Lambda_{in}^{S}({\bf k},\beta,\eta), \Lambda_{in}^{A}({\bf k},\beta,\eta)\}}$ in terms of 
${\{\Lambda_{out}^{S}({\bf k},\beta,\eta), \Lambda_{out}^{A}({\bf k},\beta,\eta)\}}$, depend solely on the description of ${g_{in}^{(\pm)}(\eta)}$ in terms of ${g_{out}^{(\pm)}(\eta)}$. From now on we introduce the correct normalization factors, namely ${N_{in/out} = (2m\omega_{in/out})^{-1/2}}$ . By means of the transformations (\ref{pe40}) and (\ref{pe31}) for ${\eta\longrightarrow+\infty}$, follows from (\ref{pe51}) and (\ref{pe52}):
\begin{eqnarray}\label{pe71}
N_{in}g_{in}^{(+)}(\eta) &\approx& N_{out}\left(\frac{N_{in}}{N_{out}}\right)\frac{\Gamma\left(1-\frac{i\omega_{in}}{\rho}\right)\Gamma\left(-\frac{i\omega_{out}}{\rho}\right)}{\Gamma\left(\frac{1\pm(\sqrt{1-4r})^{*}}{2} - \frac{i\omega_{+}}{\rho}\right)\Gamma\left(\frac{1\mp(\sqrt{1-4r})^{*}}{2} - \frac{i\omega_{+}}{\rho}\right)}e^{-i\omega_{out}\eta}\nonumber\\
&+&N_{out}\left(\frac{N_{in}}{N_{out}}\right)\frac{\Gamma\left(1-\frac{i\omega_{in}}{\rho}\right)\Gamma\left(\frac{i\omega_{out}}{\rho}\right)}{\Gamma\left(\frac{1\mp(\sqrt{1-4r})^{*}}{2} + \frac{i\omega_{-}}{\rho}\right)\Gamma\left(\frac{1\pm(\sqrt{1-4r})^{*}}{2} + \frac{i\omega_{-}}{\rho}\right)}e^{i\omega_{out}\eta}, 
\end{eqnarray}
\begin{eqnarray}\label{pe72}
N_{in}g_{in}^{(-)}(\eta) &\approx& N_{out}\left(\frac{N_{in}}{N_{out}}\right)\frac{\Gamma\left(1+\frac{i\omega_{in}}{\rho}\right)\Gamma\left(-\frac{i\omega_{out}}{\rho}\right)}{\Gamma\left(\frac{1\pm(\sqrt{1-4r})^{*}}{2} - \frac{i\omega_{-}}{\rho}\right)\Gamma\left(\frac{1\mp(\sqrt{1-4r})^{*}}{2} - \frac{i\omega_{-}}{\rho}\right)}e^{-i\omega_{out}\eta}\nonumber\\
&+&N_{out}\left(\frac{N_{in}}{N_{out}}\right)\frac{\Gamma\left(1+\frac{i\omega_{in}}{\rho}\right)\Gamma\left(\frac{i\omega_{out}}{\rho}\right)}{\Gamma\left(\frac{1\mp(\sqrt{1-4r})^{*}}{2} + \frac{i\omega_{+}}{\rho}\right)\Gamma\left(\frac{1\pm(\sqrt{1-4r})^{*}}{2} + \frac{i\omega_{+}}{\rho}\right)}e^{i\omega_{out}\eta}.  
\end{eqnarray}
Identifying the Bogolyubov coefficients from Eqs. (\ref{pe71}) and (\ref{pe72}) into (\ref{pe70}), we finally obtain the total density of created Elko spinor particles as:
\begin{eqnarray}\label{pe73}
\langle N_{\lambda}\rangle = 2\frac{\omega_{out}}{\omega_{in}}&\Bigg\{&\left\lvert\frac{\Gamma\left(1-\frac{i\omega_{in}}{\rho}\right)\Gamma\left(\frac{i\omega_{out}}{\rho}\right)}{\Gamma\left(\frac{1\mp(\sqrt{1-4r_-})^{*}}{2} + \frac{i\omega_{-}}{\rho}\right)\Gamma\left(\frac{1\pm(\sqrt{1-4r_-})^{*}}{2} + \frac{i\omega_{-}}{\rho}\right)}\right\rvert^{2}\nonumber\\
&+&\left\lvert\frac{\Gamma\left(1+\frac{i\omega_{in}}{\rho}\right)\Gamma\left(-\frac{i\omega_{out}}{\rho}\right)}{\Gamma\left(\frac{1\pm(\sqrt{1-4r_+})^{*}}{2} - \frac{i\omega_{-}}{\rho}\right)\Gamma\left(\frac{1\mp(\sqrt{1-4r_+})^{*}}{2} - \frac{i\omega_{-}}{\rho}\right)}\right\rvert^{2}\Bigg\}\,.
\end{eqnarray}
Notice that the first term must be calculated with $r_-$ and the second one with $r_+$, according to the discussion after (\ref{pe24}). The factor 2 comes from the two helicities into the sum (\ref{pe67}), namely $\beta=\{\pm,\mp \}$. This is the main result of this work. Several interesting characteristics will be studied in the next section.

\section{Comparison with scalar and Dirac fermion creation}

Now we proceed to compare result (\ref{pe73}) for the density of created Elko particles for the asymptotic metric (\ref{pe15}), namely $a(\eta)=\sqrt{A+B\tanh(\rho\eta)}$, with the well known similar results for real scalar particles \cite{davies,duncan,moradi1}:
\begin{eqnarray}\label{pe74}
\langle N_{\phi}\rangle =  2\frac{\omega_{out}}{\omega_{in}}\left\lvert\frac{\Gamma\left(1-\frac{i\omega_{in}}{\rho}\right)\Gamma\left(\frac{i\omega_{out}}{\rho}\right)}{\Gamma\left(\frac{i\omega_{-}}{\rho}\right)\Gamma\left(1 + \frac{i\omega_{-}}{\rho}\right)}\right\rvert^{2}\,.
\end{eqnarray}
The factor 2 comes from the fact that the anti-particle of a real scalar field is its own particle, being created in equal amount. Taking $r_\mp =0$ into (\ref{pe73}) we obtain twice the value of (\ref{pe74}), due to two helicities of the Elko particles.

Since the Elko is a fermionic particle, we also would like to compare its creation with that of Dirac fermions. However, an analytic expression for creation of Dirac fermionic particles for the metric (\ref{pe15}) is not known. Nevertheless an analytic expression exist for a metric of the form \cite{moradi2} $a(\eta)=C+D\tanh(\rho\eta)$. Taking into account that $\tanh(\rho\eta)$ is limited to $\pm 1$ and considering the limit $B/A\ll 1$ in our case, the metric (\ref{pe15}) can be expanded as $a(\eta)\simeq \sqrt{A} + (B/2\sqrt{A})\tanh(\rho\eta)+\dots$, thus we can compare our result (\ref{pe73}) with the corresponding for Dirac fermionic particles \cite{moradi2}:
\begin{eqnarray}\label{pe75}
\langle N_{\psi}\rangle =2\frac{N_{in}}{N^{out}}\frac{\left\lvert\Gamma\left(\frac{i\omega_{+}}{\rho} - \frac{imD}{\rho}\right)\Gamma\left(1 + \frac{i\omega_{+}}{\rho} + \frac{imD}{\rho}\right)\right\rvert^{2}}{\left\lvert\Gamma\left(1-\frac{i\omega_{-}}{\rho} + \frac{imD}{\rho}\right)\Gamma\left(-\frac{i\omega_{-}}{\rho} - \frac{imD}{\rho}\right)\right\rvert^{2} + \left\lvert\Gamma\left(\frac{i\omega_{+}}{\rho} - \frac{imD}{\rho}\right)\Gamma\left(1 + \frac{i\omega_{+}}{\rho} + \frac{imD}{\rho}\right)\right\rvert^{2}}\nonumber\\
\end{eqnarray}
with $C=\sqrt{A}$, $D=B/2\sqrt{A}$, $\omega_\pm =(\omega_{out}\pm \omega_{in})/2$ and $\omega_{in/out}=\sqrt{k^2+ m^2(C\mp D)^2}$. The normalization are given by \cite{moradi2}:
\begin{equation}
N_{in}^{out}\frac{1}{\sqrt{2m^2(C\pm D)^2+2m(C\pm D)\sqrt{k^2+m^2(C\pm D)^2}}}\,.
\end{equation}
The factor 2 comes from the fact that the number of anti-particles created is equal to the particles in the Dirac case.

Figure 2 shows the behaviour of the density of created particles as a function of its momentum for several different values of the constants. Dirac particles are in red line, Elko particles in black line and scalar particles in green line. Several interesting aspects can be seen in the graphics. For same masses, the spectrum of created Dirac particles are always greater than the Elko and scalar particles. For all kinds of particles the maximum occurs for null momenta, an indication that rest particles are created in greater quantity. Higher momenta particles demand more energy to be created, so they are suppressed. Taking $r_\mp =0$ for the Elko number density we recover exactly the spectrum of the scalar particles, with a factor 2 coming from the two kinds of Elko, namely self-conjugate and anti-self-conjugate. We also notice that greater the distance between the parameters $A$ and $B$, lesser the number of created particles (Figs. 2 (a), (b) and (c)). In Fig. 2 (d) we see the effect of the parameter $\rho$, that raises the number of created particles and enlarges the form of the spectrum (compared to Fig. 2 (b) for the same values of parameters).

In Figure 3 we compare the number of created particles as a function of its mass for some constant values of momenta. Interesting to notice that for both Dirac and scalar particles the number density of created particles goes to zero in the zero mass limit for every momenta, while for Elko particles the number density increases in the null mass limit. A second maximum occurs for Elko of small masses and momenta (Figs. 3 (a) and (b)), in agreement to the maximum that also there exist for scalar and Dirac particles.  In the specific case of null mass and null momenta the number density diverges for Elko particles. Such behaviour is showed in Figure 4, for a 3D plotting. The presence of a second maximum is also evident from Fig. 4 (a). Although such result point to a infinite amount of created Elko particles with zero mass at rest, the very definition of the Elko field assures it must be massive. Thus we cannot have zero mass Elko particles being created, since that such spinor is not defined at this limit.

As a last analysis we compare in Figure 5 the number density of created particles just for Elko and Dirac, for slightly different masses as a function of its momenta. Notice that the creation of Elko particles is greater than Dirac particles when its mass is lesser than the Dirac one (normalized to one). This has a interesting interpretation if we consider the Elko as the candidate to dark matter in the universe. Lighter Elko particles are created in larger quantities than Dirac fermionic particles. Such difference could be responsible for the presence of about 5\% of baryonic matter and 25\% of dark matter particles in the actual universe. Being more light than the baryonic particles, the Elko particles are more difficult to be detected into accelerators for instance.

\section{Concluding remarks}

In this paper we have analysed the gravitational particle creation of the Elko spinor field for a metric scale factor of the form $a(\eta)=\sqrt{A+B\tanh \rho \eta}$. Such model admits exact analytic solutions for real scalar particles in general case and also for Dirac particles in the limit $B/A\ll 1$, thus we have conditions to compare the spectrum of creation of Elko particles with real scalar and fermionic Dirac particles.

The differential equation that governs the time mode function was obtained, Eq. (\ref{pe14}), being exactly a generalization of real scalar particles. Analytic solutions of such equation were obtained for the aforementioned scale factor after using an approximation that permits to write the solutions by means of generalized hypergeometric functions, Eq. (\ref{pe24}). Finally, the number density of created Elko particles was obtained, Eq. (\ref{pe73}). All the above results are generalization of the corresponding to real scalar particles. Although being a fermion, the time dependent part of the Elko spinor field follows an evolution similar to scalar particles. Even so, in which concerns the spectrum of created particles, several interesting properties are present, as follows.

For particles of same mass we have obtained that the number density of created Elko particles is greater than scalar particles and lesser than Dirac particles. For the three kinds of particles the maximum of creation occurs for null momenta, which is reasonable, since production of high kinetic energy particles needs much more gravitational field variation. 

The spectrum of density of created particles as a function of the its masses were also analysed for some constant values of the momenta. Dirac and scalar particles have a maximum production for a non null mass and goes to zero in the null and high mass limit, similar to a thermal spectrum. Elko particles presents a peculiar behaviour. In the zero mass limit the creation is maximum, going to infinity for null momenta. Another consequence is that lighter Elko particles can be created in more quantity than Dirac particles. By considering the Elko particles as candidate to dark matter in the universe, we interpreted that there exist much more light Elko particles created gravitationally than baryonic fermionic particles. Such result is confirmed by observations, since that there are about 25\% of dark matter against about 5\% of baryonic matter, according to standard model of cosmology.

\begin{figure*}[t]
\begin{center}
\centering
\includegraphics[width=0.45\textwidth]{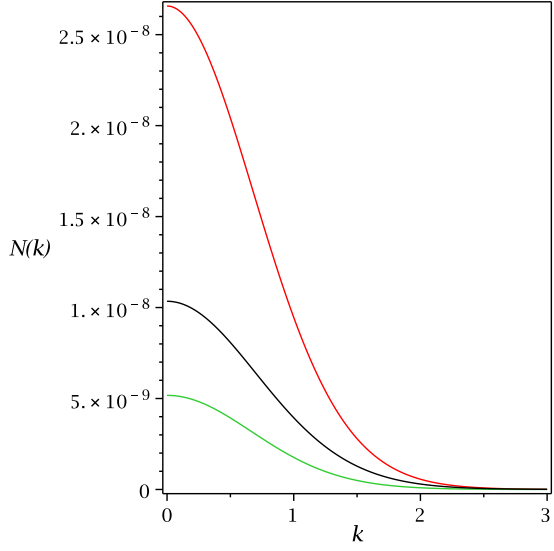}
\hspace{1.3cm}
\includegraphics[width=0.45\textwidth]{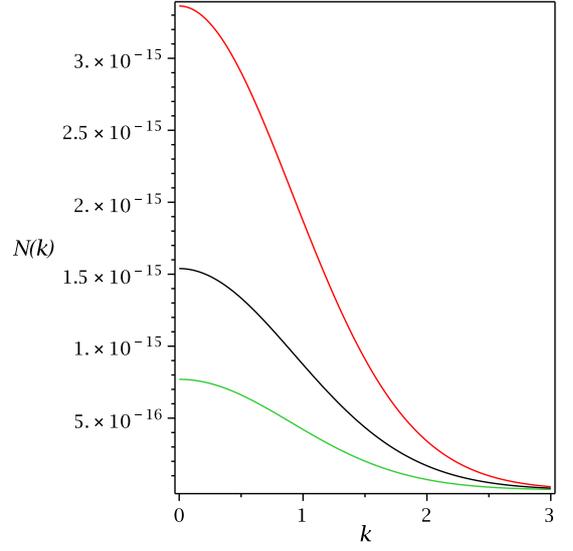}\\
\hspace{1.9cm}(a)\hspace{8.5cm}(b)\\
\vspace{0.7cm}
\includegraphics[width=0.45\textwidth]{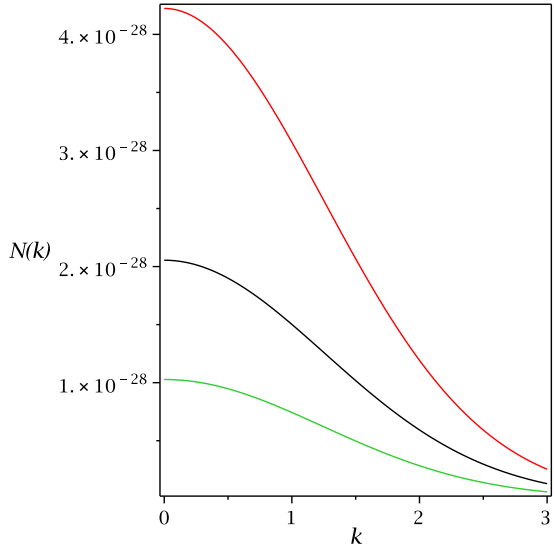}
\hspace{1.3cm}
\includegraphics[width=0.45\textwidth]{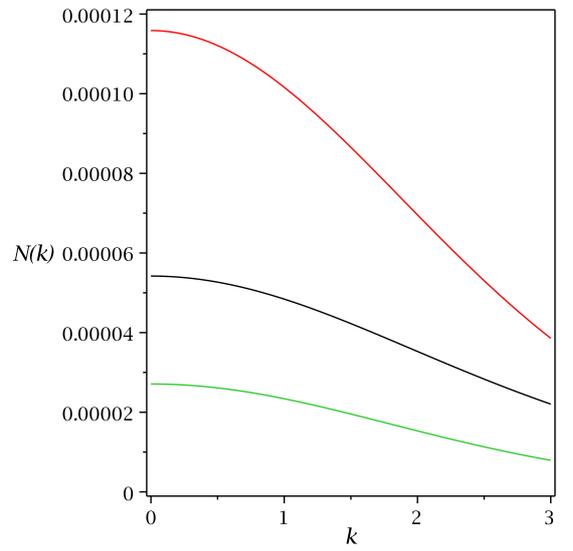}\\
\hspace{1.9cm}(c)\hspace{8.5cm}(d)\\
\end{center}
\caption{Graphical analysis for $N_\psi$ (Dirac - red line), $N_\lambda$ (Elko - black line) and $N_\phi$ (Scalar - green line) as a function of the momenta for different values of parameters:  (a) $A=10$, $B=1$, $m=1$, $\rho=1$; (b) $A=30$, $B=1$, $m=1$, $\rho=1$; (c) $A=100$, $B=1$, $m=1$, $\rho=1$; (d) $A=30$, $B=1$, $m=1$, $\rho=5$. }
\end{figure*}\label{fig2}

\begin{figure*}[t]
\begin{center}
\centering
\includegraphics[width=0.45\textwidth]{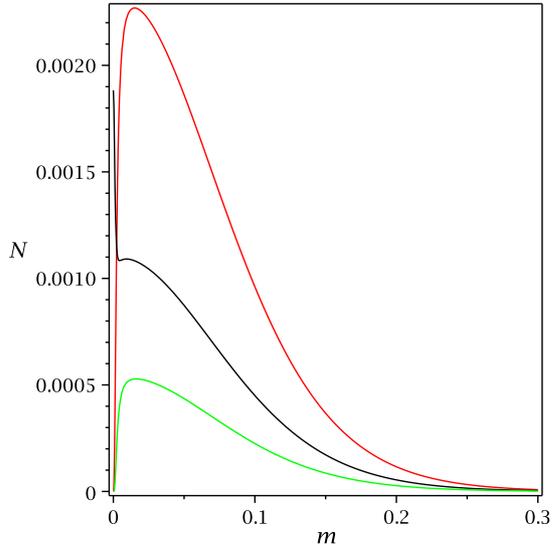}
\hspace{1.3cm}
\includegraphics[width=0.45\textwidth]{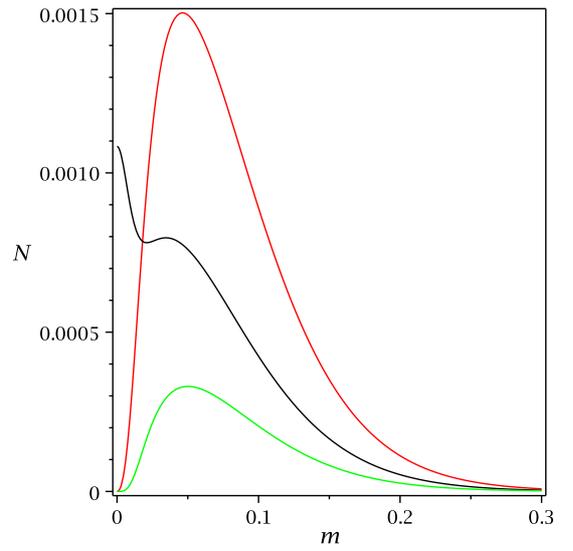}\\
\hspace{1.7cm}(a)\hspace{8.5cm}(b)\\
\vspace{0.7cm}
\includegraphics[width=0.45\textwidth]{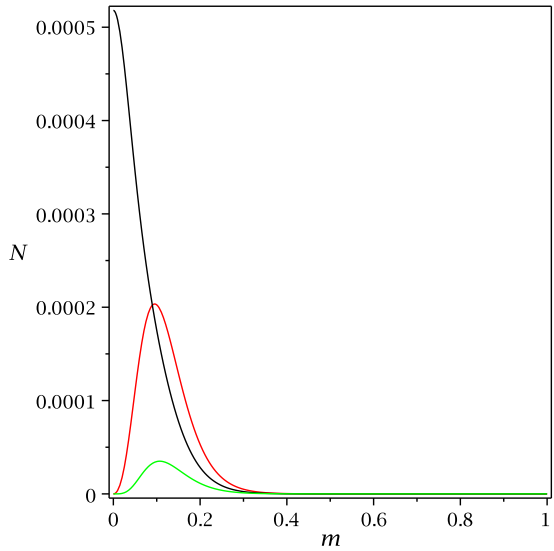}
\hspace{1.3cm}
\includegraphics[width=0.45\textwidth]{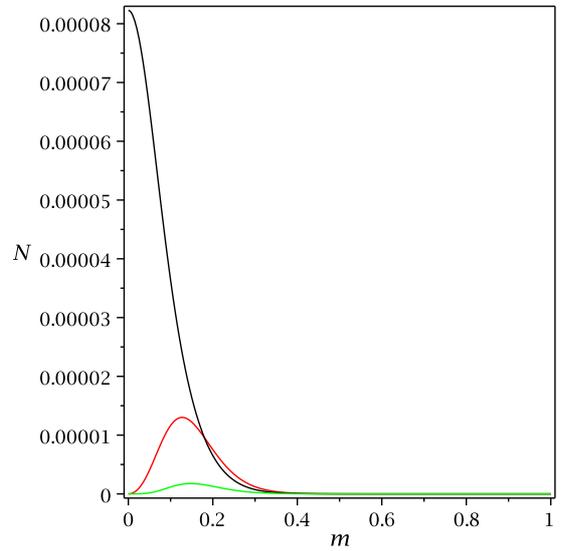}\\
\hspace{1.7cm}(c)\hspace{8.5cm}(d)
\end{center}
\caption{Graphical analysis for $N_\psi$ (Dirac - red line), $N_\lambda$ (Elko - black line) and $N_\phi$ (Scalar - green line) as a function of the mass for a fixed momenta $k$ and parameters $A=30$, $B=1$, $\rho=1$: (a) $k=0.01$; (b) $k=0.1$; (c) $k=0.5$; (d) $k=1.0$.}\label{fig3}
\end{figure*}

\begin{figure*}[t]
\begin{center}
\centering
\includegraphics[width=0.47\textwidth]{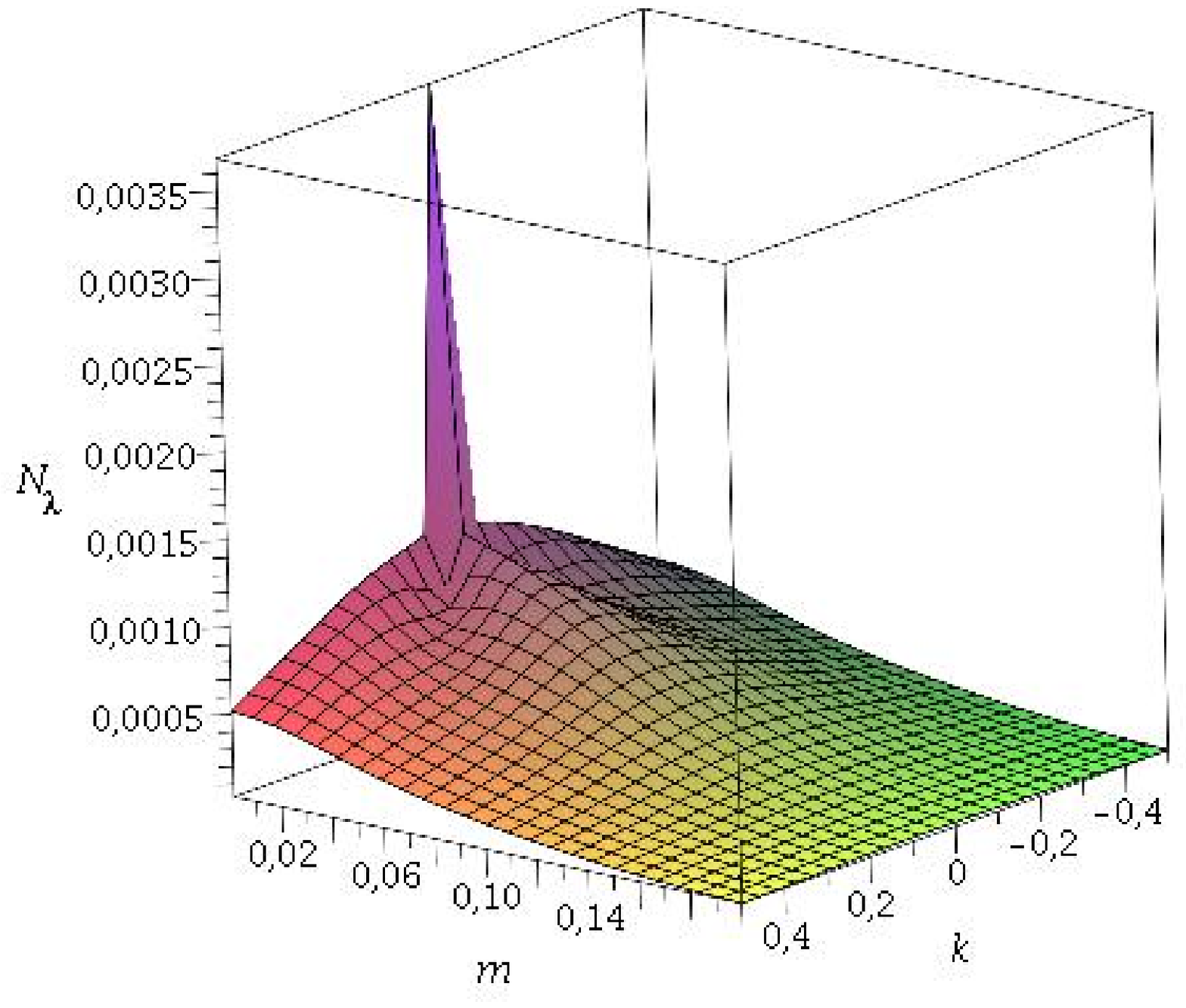}
\includegraphics[width=0.52\textwidth]{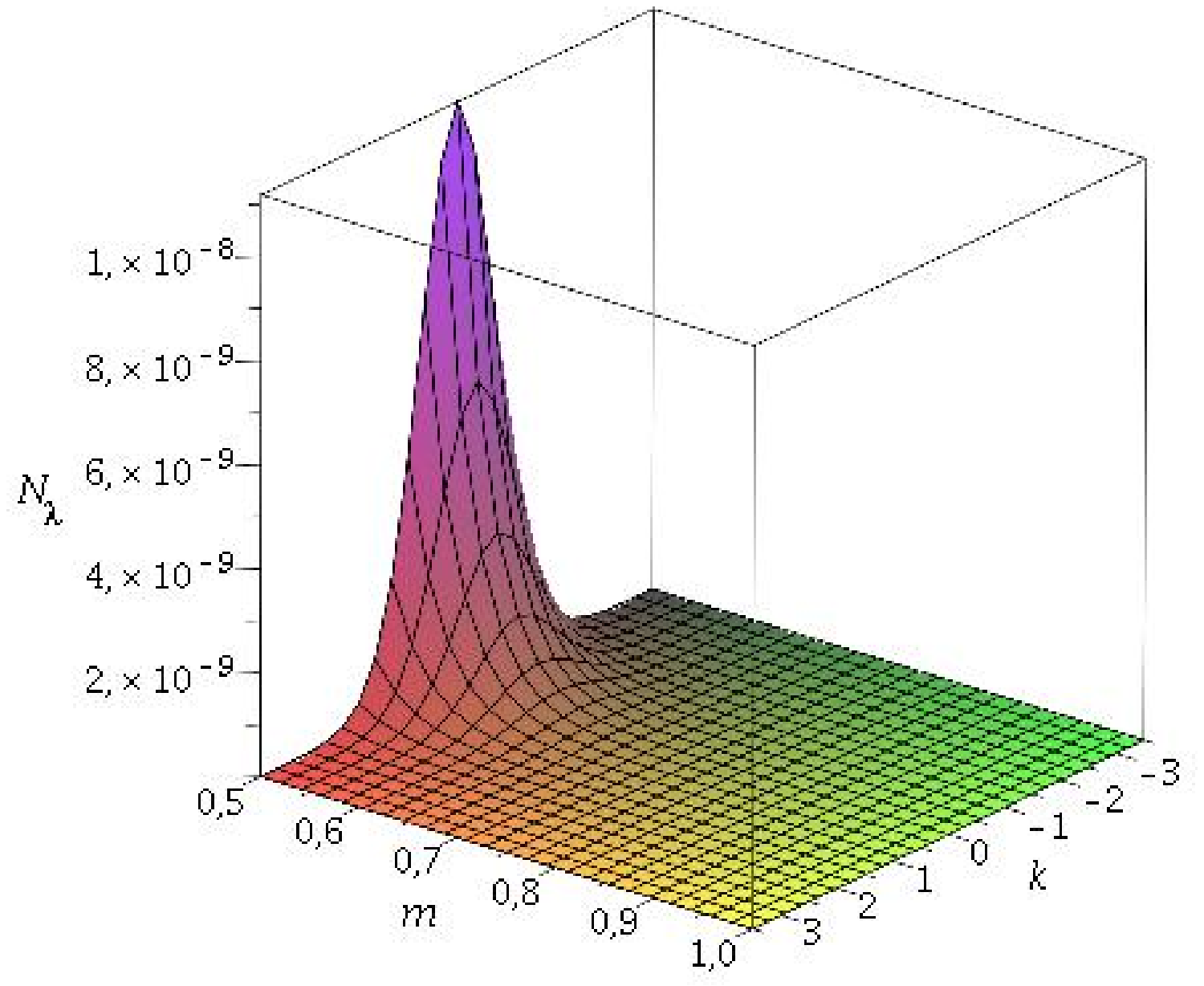}\\
\hspace{1.9cm}(a)\hspace{8.7cm}(b)
\end{center}
\caption{3D plot for the number density of created Elko particles as a function of its mass and momenta: (a) Region of small mass and momenta; (b) Region of greater mass and momenta.}\label{fig4}
\vspace{0.8cm}
\begin{center}
\centering
\includegraphics[width=0.45\textwidth]{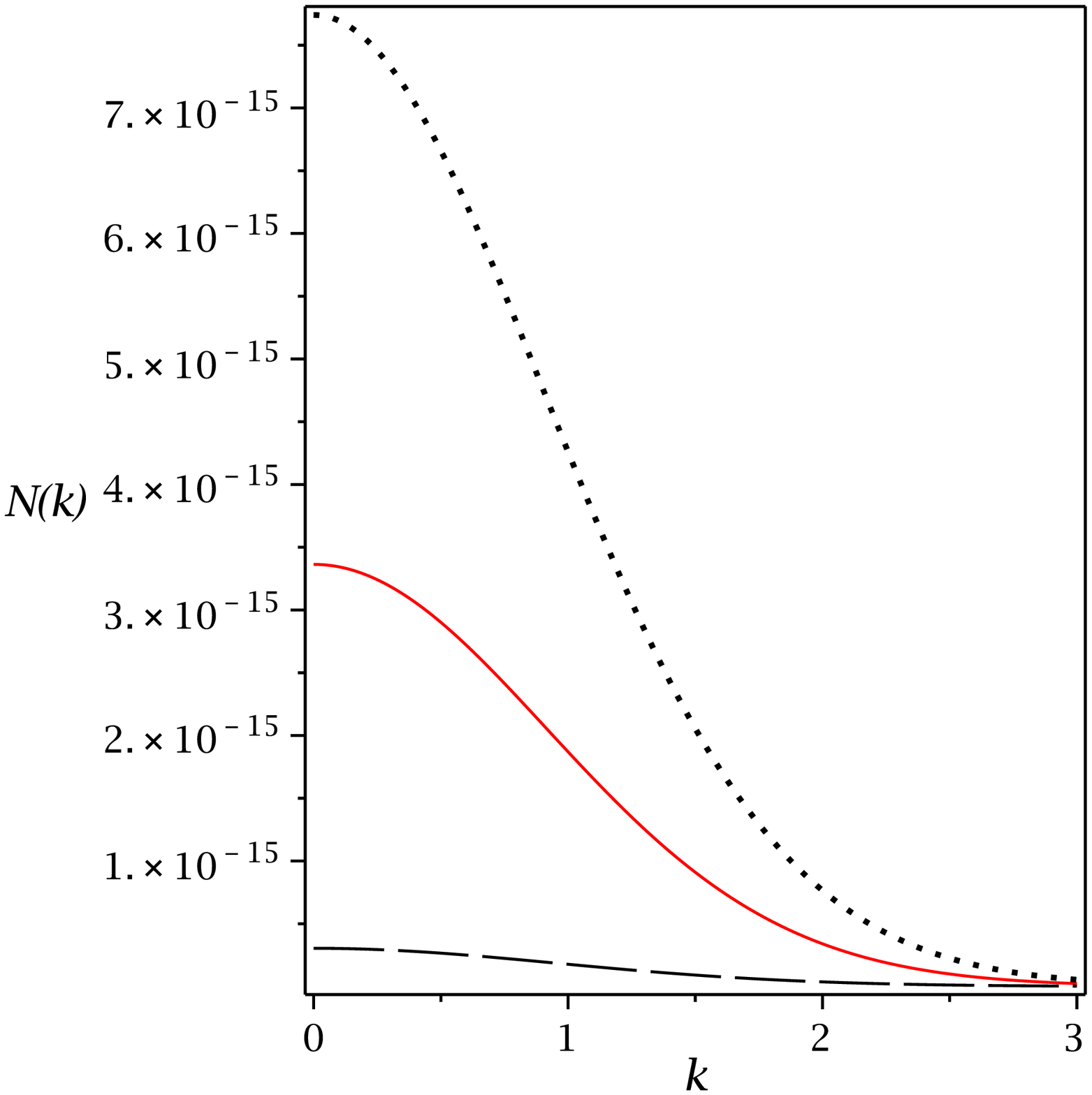}
\hspace{1.3cm}
\includegraphics[width=0.45\textwidth]{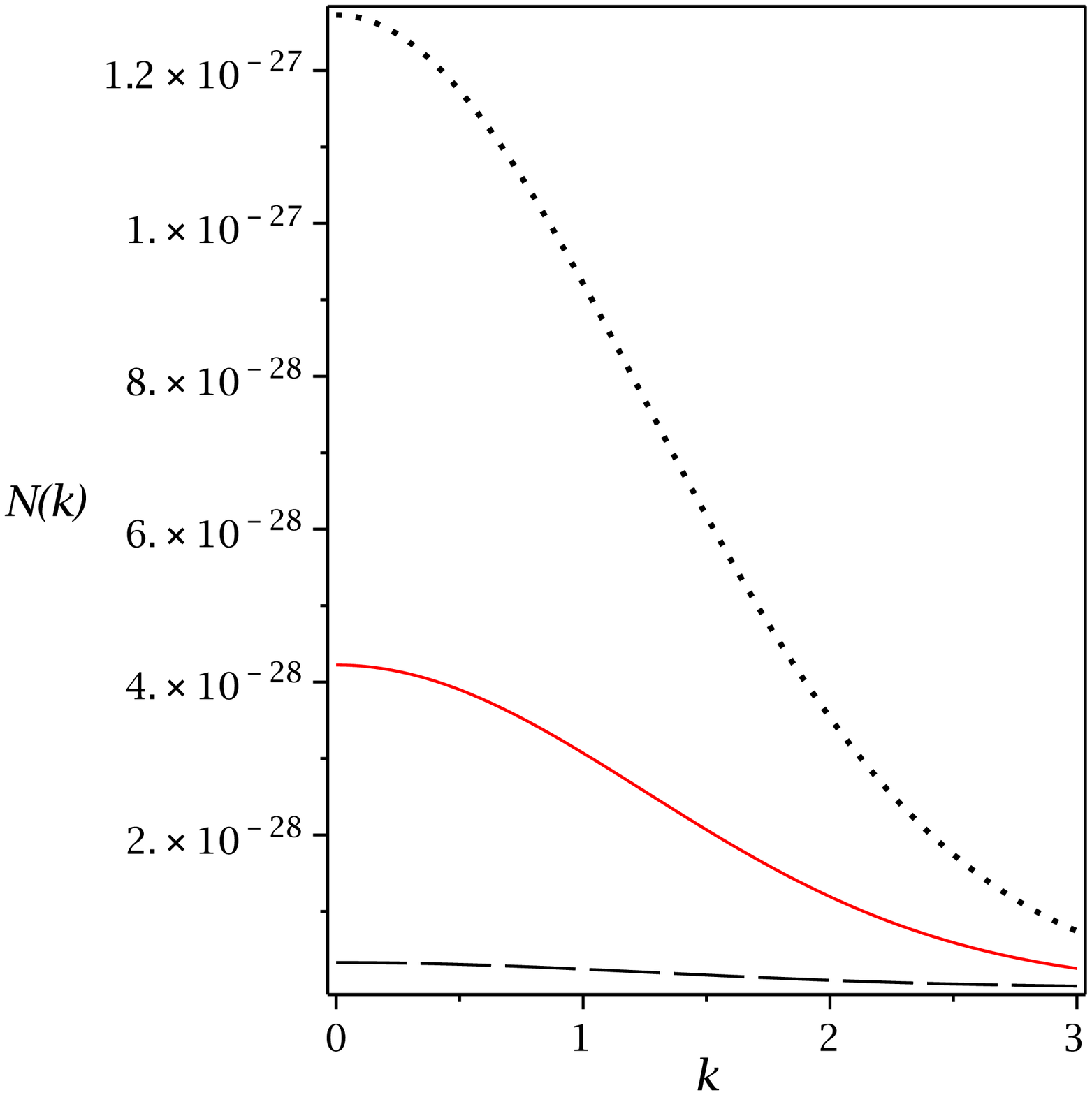}\\
\hspace{1.9cm}(a)\hspace{8.7cm}(b)
\end{center}
\caption{Graphical analysis for $N_\psi$ (Dirac) and $N_\lambda$ (Elko) as a function of the momenta for different masses, with the following values of parameters ($\rho=1$):  (a) $A=30$, $B=1$,  $m_{Dirac}=1$ (red line), $m_{elko}=1.05 m_{Dirac}$ (black dashed) and  $m_{elko}=0.95 m_{Dirac}$ (black dot); (b) $A=100$, $B=1$,  $m_{Dirac}=1$ (red line), $m_{elko}=1.03 m_{Dirac}$ (black dashed) and  $m_{elko}=0.97 m_{Dirac}$ (black dot).}\label{fig5}
\end{figure*}

\newpage

\section*{APPENDIX}

\subsection{Elko spinor field}
The Elko spinors $\lambda({\bf k})$ are constructed by imposing they are eigenspinors of the charge conjugation operator $C$ \cite{AHL1,AHL2,AHL3}:
 \begin{eqnarray}\label{cee1}
 C\lambda({\bf k}) = \pm\lambda({\bf k}) \Rightarrow \begin{cases}
                                                      C\lambda^{S}({\bf k}) = +1.\lambda^{S}({\bf k})\\
                                                      C\lambda^{A}({\bf k}) = -1.\lambda^{A}({\bf k})
                                                     \end{cases}
\end{eqnarray}
with ${\lambda^{S}({\bf k})}$ and ${\lambda^{A}({\bf k})}$ standing for self-conjugate and anti-self-conjugate spinors, respectively, with eingenvalues ${+1}$ and ${-1}$. The operator $C$ is given by, in the Weyl representation:
\begin{equation}\label{cee2}
C \stackrel{Weyl}{=} \begin{pmatrix}
                        0_{2\times2} & i\Theta\\ 
                        -i\Theta     & 0_{2\times2}   
                       \end{pmatrix}K,                    
\end{equation}
where ${\Theta =       \begin{pmatrix}
                        0 & -1\\ 
                        1 & 0  
                       \end{pmatrix}}$ is the Wigner time reverse operator and $K$ just takes the complex conjugation of the fields on the right.
                      
 With the Dirac $\gamma$ matrices given by ${\gamma^{0} = \begin{pmatrix}
                              0_{2\times2} & I_{2\times2}\\ 
                              I_{2\times2} & 0_{2\times2}  
                            \end{pmatrix}}$
e            ${\gamma^{i} = \begin{pmatrix}
                              0_{2\times2} & -\sigma^{i}_{2\times2}\\ 
                              \sigma^{i}_{2\times2} & 0_{2\times2}  
                            \end{pmatrix}}$, where ${\sigma^{i} = \sigma_{i}}$ are the Pauli matrices, we have four spinors satisfying (\ref{cee2}):
                            \begin{eqnarray}\label{cee3}
&&\lambda^{S}_{\{+,-\}}({\bf 0}) = \begin{pmatrix} 
                  +\sigma_{2}[\phi_{L}^{-}({\bf 0})]^{*} \\ 
                        \phi_{L}^{-}({\bf 0})   
                       \end{pmatrix} \longmapsto \begin{pmatrix} 
                                   \Uparrow\\ 
                                   \Downarrow
                       \end{pmatrix},\nonumber\\
&&\lambda^{S}_{\{-,+\}}({\bf 0}) = \begin{pmatrix} 
                  +\sigma_{2}[\phi_{L}^{+}({\bf 0})]^{*} \\ 
                        \phi_{L}^{+}({\bf 0})   
                       \end{pmatrix} \longmapsto \begin{pmatrix} 
                                   \Downarrow\\ 
                                   \Uparrow
                       \end{pmatrix},\nonumber\\
&&\lambda^{A}_{\{+,-\}}({\bf 0}) = \begin{pmatrix} 
                  -\sigma_{2}[\phi_{L}^{-}({\bf 0})]^{*} \\ 
                        \phi_{L}^{-}({\bf 0})   
                       \end{pmatrix} \longmapsto \begin{pmatrix} 
                                   \Uparrow\\ 
                                   \Downarrow
                       \end{pmatrix},\nonumber\\
&&\lambda^{A}_{\{-,+\}}({\bf 0}) = \begin{pmatrix} 
                  -\sigma_{2}[\phi_{L}^{+}({\bf 0})]^{*} \\ 
                        \phi_{L}^{+}({\bf 0})   
                       \end{pmatrix} \longmapsto \begin{pmatrix} 
                                   \Downarrow\\ 
                                   \Uparrow
                       \end{pmatrix},
\end{eqnarray}
where ${\beta =}$ (${\{+,-\}}$,${\{-,+\}}$) stands for the helicities, pictorially represented on the right, and
${\phi_{L}^{\pm}({\bf 0})}$ are the left-handed Weyl at rest:
\begin{eqnarray}\label{cee4}
&&\phi_{L}^{+}({\bf 0}) = \sqrt{m}\begin{pmatrix} 
                                   cos(\theta/2)e^{-i\varphi/2}\\ 
                                   sen(\theta/2)e^{+i\varphi/2}
                                  \end{pmatrix},\nonumber\\
&&\phi_{L}^{-}({\bf 0}) = \sqrt{m}\begin{pmatrix} 
                                   -sen(\theta/2)e^{-i\varphi/2}\\ 
                                   cos(\theta/2)e^{+i\varphi/2}
                                  \end{pmatrix},
\end{eqnarray}
and ${m}$ represents its mass. In the null mass limit the Elko spinor is not defined. For an arbitrary moment we just apply a Lorentz boost:
\begin{eqnarray}\label{cee5}
&&\lambda^{S}_{\{\pm,\mp\}}({\bf k}) = \sqrt{\frac{E + m}{2m}}\left(1 \pm \frac{k}{E+m}\right)\lambda^{S}_{\{\pm,\mp\}}({\bf 0}),\nonumber\\
&&\lambda^{A}_{\{\pm,\mp\}}({\bf k}) = \sqrt{\frac{E + m}{2m}}\left(1 \pm \frac{k}{E+m}\right)\lambda^{A}_{\{\pm,\mp\}}({\bf 0}),
\end{eqnarray}
with ${E = (k^{2} + m^{2})^{1/2}}$ the energy.

The duals $\stackrel{\neg}{\lambda}_{\beta}({\bf k})$ are constructed such that  ${\stackrel{\neg}{\lambda}_{\beta}({\bf k})\lambda_{\beta}({\bf k})}$ are Lorentz invariants, so they are defined as:
\begin{equation}\label{cee6}
\stackrel{\neg}{\lambda}_{\{\mp,\pm\}}^{S/A}({\bf k}) = \pm i[\lambda_{\{\pm,\mp\}}^{S/A}({\bf k})]^{\dag}\gamma^{0},
\end{equation}
With (\ref{cee5}) and (\ref{cee6}) we obtain the following properties:
\begin{eqnarray}\label{cee7}
&&\stackrel{\neg}{\lambda}_{\beta}({\bf k})^{S}\lambda_{\beta^{'}}({\bf k})^{S} = 2m\delta_{\beta\beta^{'}},\nonumber\\
&&\stackrel{\neg}{\lambda}_{\beta}({\bf k})^{A}\lambda_{\beta^{'}}({\bf k})^{A} = -2m\delta_{\beta\beta^{'}}
\end{eqnarray}

In Ref. \cite{AHL4} a new dual was proposed leading to important consequences in their quantum fields, with some additional support given in \cite{RJ} and \cite{WT}. The important fact to our study is that such a new dual preserves the relations (\ref{cee7}).

\subsection{Riemann-Papperitz equation}
The Riemann-Papperitz differential equation for ${w = w(z)}$ is given by \cite{25}:
\begin{eqnarray}\label{A10}
&&\frac{d^{2}w}{dz^{2}} + \left[\frac{1-\alpha-\alpha^{'}}{z-z_{1}} + \frac{1-\beta-\beta^{'}}{z-z_{2}} + \frac{1-\gamma-\gamma^{'}}{z-z_{3}}\right]\frac{dw}{dz}\nonumber\\
&&+\left[\frac{\alpha\alpha^{'}(z_{1}-z_{2})(z_{1}-z_{3})}{z-z_{1}} + \frac{\beta\beta^{'}(z_{2}-z_{3})(z_{2}-z_{1})}{z-z_{2}} + \frac{\gamma\gamma^{'}(z_{3}-z_{1})(z_{3}-z_{2})}{z-z_{3}}\right]\nonumber\\
&&\times\frac{w}{(z-z_{1})(z-z_{2})(z-z_{3})} = 0,
\end{eqnarray}
where ${z_{1}}$, ${z_{2}}$ and ${z_{3}}$ are the poles of the equation and ${\alpha}$, ${\alpha^{'}}$, ${\beta}$, ${\beta^{'}}$, ${\gamma}$ 
and ${\gamma^{'}}$ de Riemann coefficients that satisfies: 
\begin{equation}\label{A11}
\alpha + \alpha^{'} + \beta + \beta^{'} + \gamma + \gamma^{'} - 1 = 0. 
\end{equation}
Choosing the poles as ${z_{1} = 0}$, ${z_{2} = 1}$ and ${z_{3} = \infty}$, (\ref{A10}) is:
\begin{eqnarray}\label{A12}
&&\frac{d^{2}w}{dz^{2}} + \left[\frac{1-\beta-\beta^{'}}{z-1} + \frac{1-\alpha-\alpha^{'}}{z}\right]\frac{dw}{dz}+\left[\frac{\beta\beta^{'}}{z-1} - \frac{\alpha\alpha^{'}}{z} + \gamma\gamma^{'}\right]\frac{w}{z(z-1)} = 0,
\end{eqnarray}
whose solution is: 
\begin{equation}\label{A13}
w(z) = P\begin{Bmatrix} 
z_{1} & z_{2} & z_{3} \\ 
\alpha & \beta & \gamma & z\\ 
\alpha^{'} & \beta^{'} & \gamma^{'}  
\end{Bmatrix} = P\begin{Bmatrix} 
0 & 1 & \infty \\ 
\alpha & \beta & \gamma & z\\ 
\alpha^{'} & \beta^{'} & \gamma^{'}  
\end{Bmatrix},
\end{equation}
represented by the ${P}\{\}$ Riemann symbol, which admits the following transformation for the poles:
\begin{equation}\label{A14}
P\begin{Bmatrix} 
0 & 1 & \infty \\ 
\delta + \kappa & \varepsilon + l & \zeta - \kappa - l & z\\ 
\delta^{'} + \kappa & \varepsilon^{'} + l & \zeta^{'} - \kappa - l  
\end{Bmatrix} = z^{\kappa}(1-z)^{l}P\begin{Bmatrix} 
0 & 1 & \infty \\ 
\delta& \varepsilon & \zeta & z\\ 
\delta^{'} & \varepsilon^{'}& \zeta^{'}  
\end{Bmatrix}
\end{equation}
\\
Now, choosing the coefficients of (\ref{A12}) as
\begin{eqnarray}\label{A16}
&&\begin{cases}
   1-\alpha-\alpha^{'} = 1\\
   \alpha\alpha^{'} = \frac{\omega_{in}^{2}}{4\rho^{2}}
  \end{cases} \Rightarrow \alpha = -\alpha^{'} = i\frac{\omega_{in}}{2\rho}, 
\end{eqnarray}
\begin{eqnarray}\label{A17}
&&\begin{cases}
   1-\beta-\beta^{'} = 1\\
   \beta\beta^{'} = \frac{\omega_{out}^{2}}{4\rho^{2}}
  \end{cases} \Rightarrow \beta = -\beta^{'} = i\frac{\omega_{out}}{2\rho}, 
\end{eqnarray}
\begin{eqnarray}\label{A18}
&&\begin{cases}
   1-\gamma-\gamma^{'} = 0\\
   \gamma\gamma^{'} = r
  \end{cases} \Rightarrow \gamma = \frac{1\pm\sqrt{1-4r}}{2}, \gamma^{'} = \frac{1\mp\sqrt{1-4r}}{2},  
\end{eqnarray}
where ${z \rightarrow \xi}$, ${w(z) \rightarrow \tilde{g}(\xi)}$ and ${r = \frac{B}{8(A+B)}\left(\frac{3c_{2}B}{2(A+B)} - \frac{4c_{1}(X_{j} + iY_{j})}{\rho}\right)}$.
We obtain for ${\tilde{g}(\xi)}$:
\begin{eqnarray}\label{A15}
&&\Bigg[\frac{d^{2}}{d\xi^{2}} + \left(\frac{1}{\xi-1} + \frac{1}{\xi}\right)\frac{d}{d\xi} + \left(\frac{\omega_{out}^{2}}{4\rho^{2}}\frac{1}{\xi-1} - \frac{\omega_{in}^{2}}{4\rho^{2}}\frac{1}{\xi}\right)\cdot\frac{1}{\xi(\xi-1)}\nonumber\\
&&+\frac{B}{8(A+B)}\left(\frac{3c_{2}B}{2(A+B)} - \frac{4c_{1}(X_{j} + iY_{j})}{\rho}\right)\cdot\frac{1}{\xi(\xi-1)}\Bigg]\tilde{g}(\xi) = 0,
\end{eqnarray}
which is exactly (\ref{pe23}). The solution comes from (\ref{A13}):
\begin{equation}\label{A19}
\tilde{g}(\xi) = P\begin{Bmatrix} 
0 & 1 & \infty \\ 
i\frac{\omega_{in}}{2\rho} & i\frac{\omega_{out}}{2\rho} & \frac{1\pm\sqrt{1-4r}}{2} & \xi\\ 
-i\frac{\omega_{in}}{2\rho} & -i\frac{\omega_{out}}{2\rho} & \frac{1\mp\sqrt{1-4r}}{2}  
\end{Bmatrix}.
\end{equation}
Using the transformation (\ref{A14}) into (\ref{A19}):
\begin{equation}\nonumber
\tilde{g}(\xi) = P\begin{Bmatrix} 
0 & 1 & \infty \\ 
i\frac{\omega_{in}}{2\rho} & i\frac{\omega_{out}}{2\rho} & \frac{1\pm\sqrt{1-4r}}{2} & \xi\\ 
-i\frac{\omega_{in}}{2\rho} & -i\frac{\omega_{out}}{2\rho} & \frac{1\mp\sqrt{1-4r}}{2}  
\end{Bmatrix} = P\begin{Bmatrix} 
0 & 1 & \infty \\ 
\delta + \kappa & \varepsilon + l & \zeta - \kappa - l & \xi\\ 
\delta^{'} + \kappa & \varepsilon^{'} + l & \zeta^{'} - \kappa - l  
\end{Bmatrix},\nonumber
\end{equation}
and taking ${\kappa = \frac{i\omega_{in}}{2\rho}}$ and ${l = \frac{i\omega_{out}}{2\rho}}$, we have:
\begin{eqnarray}\nonumber
\begin{cases}
 \delta = 0\\
 \varepsilon = 0\\
 \zeta = \frac{1\pm\sqrt{1-4r}}{2} + \frac{i\omega_{+}}{\rho}\\
 \delta^{'} = -\frac{i\omega_{in}}{\rho}\\
 \varepsilon^{'} = -\frac{i\omega_{out}}{\rho}\\
 \zeta^{'} = \frac{1\mp\sqrt{1-4r}}{2} + \frac{i\omega_{+}}{\rho},
\end{cases} 
\end{eqnarray}
with ${\omega_{\pm} = \frac{1}{2}(\omega_{out} \pm \omega_{in})}$.\\

Finally, ${\tilde{g}(\xi)}$ from (\ref{A19}) and (\ref{A14}) can be written as:
\begin{equation}\label{A20}
\tilde{g}(\xi) = \xi^{\frac{i\omega_{in}}{2\rho}}(1-\xi)^{\frac{i\omega_{out}}{2\rho}}P\begin{Bmatrix} 
0 & 1 & \infty \\ 
0 & 0 & \frac{1\pm\sqrt{1-4r}}{2} + \frac{i\omega_{+}}{\rho} & \xi\\ 
-\frac{i\omega_{in}}{\rho} & -\frac{i\omega_{out}}{\rho} & \frac{1\mp\sqrt{1-4r}}{2} + \frac{i\omega_{+}}{\rho}  
\end{Bmatrix}
\end{equation}
which is just the solution (\ref{pe24}) for our problem.

Now we can write the above solution in terms of the standard hypergeometric function \cite{25}:
\begin{equation}\label{pe26}
_{2}F_{1}(\epsilon,\upsilon;\varpi;z) = P\begin{Bmatrix} 
0 & 1 & \infty \\ 
0 & 0 & \epsilon & z\\ 
1-\varpi & \varpi-\epsilon-\upsilon & \upsilon  
\end{Bmatrix},
\end{equation}
where $_{2}F_{1}(\epsilon,\upsilon;\varpi;z)$ is defined by the series:
\begin{equation}\label{pe27}
_{2}F_{1}(\epsilon,\upsilon;\varpi;z) = 1 + \frac{\epsilon\cdot \upsilon}{\varpi\cdot 1}z + \frac{\epsilon(\epsilon+1)\upsilon(\upsilon+1)}{\varpi(\varpi+1)\cdot 1\cdot 2}z^{2} +...,
\end{equation}
and satisfies the relation:
\begin{equation}\label{pe31}
_{2}F_{1}(\epsilon,\upsilon;\varpi;z) = {(1-z)^{\varpi-\epsilon-\upsilon}} _2F_{1}(\varpi-\epsilon,\varpi-\upsilon;\varpi;z).
\end{equation}

Comparing (\ref{A20}) with (\ref{pe26}) we have:
\begin{eqnarray}\nonumber
\begin{cases}
\epsilon = \frac{1\pm\sqrt{1-4r}}{2} + \frac{i\omega_{+}}{\rho}\\
\upsilon = \frac{1\mp\sqrt{1-4r}}{2} + \frac{i\omega_{+}}{\rho}\\
\varpi = 1 + \frac{i\omega_{in}}{\rho}\\
z = \xi
\end{cases},
\end{eqnarray}
and ${\tilde{g}(\xi)}$ can be write as a hypergeometric:
\begin{eqnarray}\label{pe28}
\tilde{g}(\xi) &=& \xi^{\frac{i\omega_{in}}{2\rho}}(1-\xi)^{\frac{i\omega_{out}}{2\rho}}\nonumber\\
               &\times&_{2}F_{1}\left(\frac{1\pm\sqrt{1-4r}}{2} + \frac{i\omega_{+}}{\rho}, \frac{1\mp\sqrt{1-4r}}{2} + \frac{i\omega_{+}}{\rho}; 1 + \frac{i\omega_{in}}{\rho};\xi\right).
\end{eqnarray}
In order to write the solution for ${g(\eta)}$ we use
\begin{eqnarray}\label{pe29}
\begin{cases}
\xi^{\frac{i\omega_{in}}{2\rho}} = \left(\frac{1+\tanh(\rho\eta)}{2}\right)^{\frac{i\omega_{in}}{2\rho}} = \left(\frac{\e^{\rho\eta}}{2\cosh(\rho\eta)}\right)^{\frac{i\omega_{in}}{2\rho}} = \e^{\frac{i\omega_{in}}{2}\eta}\e^{-\frac{i\omega_{in}}{2\rho}\ln(2\cosh(\rho\eta))},\\
(1-\xi)^{\frac{i\omega_{out}}{2\rho}} = \left(\frac{1-\tanh(\rho\eta)}{2}\right)^{\frac{i\omega_{out}}{2\rho}} = \e^{\frac{-i\omega_{out}}{2}\eta}\e^{-\frac{i\omega_{out}}{2\rho}\ln(2\cosh(\rho\eta))}.                                 
\end{cases} 
\end{eqnarray}
Substituting into (\ref{pe28}):
\begin{eqnarray}\label{pe30}
g(\eta) &=& \exp\left[-i\left(\omega_{-}\eta + \frac{\omega_{+}}{\rho}\ln(2\cosh(\rho\eta))\right)\right]\\\nonumber
               &\times&_{2}F_{1}\left(\frac{1\pm\sqrt{1-4r}}{2} + \frac{i\omega_{+}}{\rho}, \frac{1\mp\sqrt{1-4r}}{2} + \frac{i\omega_{+}}{\rho}; 1 + \frac{i\omega_{in}}{\rho};\frac{1+\tanh(\rho\eta)}{2}\right).\nonumber 
\end{eqnarray}
With equation ({\ref{pe30}}) we will obtain ${g_{in}(\eta)}$ e ${g_{out}(\eta)}$.

Now, by using (\ref{pe31}) and
\begin{eqnarray}\nonumber
\begin{cases}
\varpi-\epsilon-\upsilon = -\frac{i\omega_{out}}{\rho}\\
\varpi-\epsilon = \frac{1\mp\sqrt{1-4r}}{2} - \frac{i\omega_{-}}{\rho}\\
\varpi-\upsilon = \frac{1\pm\sqrt{1-4r}}{2} - \frac{i\omega_{-}}{\rho}
\end{cases},
\end{eqnarray}
we also have for $\tilde{g}(\xi)$
\begin{eqnarray}\label{pe33}
\tilde{g}(\xi) = \xi^{\frac{i\omega_{in}}{2\rho}}\, _{2}F_{1}\left(\frac{1\mp\sqrt{1-4r}}{2} - \frac{i\omega_{-}}{\rho}, \frac{1\pm\sqrt{1-4r}}{2} - \frac{i\omega_{-}}{\rho};1+\frac{i\omega_{in}}{\rho};\xi\right),
\end{eqnarray}
and for ${g(\eta)}$ (using (\ref{pe29})):
\begin{eqnarray}\label{pe34}
g(\eta) &=& \exp\left[i\left(\frac{\omega_{in}}{2}\eta - \frac{\omega_{in}}{2\rho}\ln(2\cosh(\rho\eta))\right)\right]\\\nonumber
               &\times&_{2}F_{1}\left(\frac{1\mp\sqrt{1-4r}}{2} - \frac{i\omega_{-}}{\rho}, \frac{1\pm\sqrt{1-4r}}{2} - \frac{i\omega_{-}}{\rho};1+\frac{i\omega_{in}}{\rho};\frac{1+\tanh(\rho\eta)}{2}\right).\nonumber 
\end{eqnarray}
In the limit $\eta\to -\infty$ we obtain:
\begin{eqnarray}\label{pe37}
g(\eta\to -\infty) &=& \exp\left[i\left(\omega_{+}\eta + \frac{\omega_{-}}{\rho}\ln(2\cosh(\rho\eta))\right)\right]\\\nonumber
                                         &\times&_{2}F_{1}\left(\frac{1\mp\sqrt{1-4r}}{2} - \frac{i\omega_{-}}{\rho}, \frac{1\pm\sqrt{1-4r}}{2} - \frac{i\omega_{-}}{\rho};1+\frac{i\omega_{in}}{\rho};\frac{1+\tanh(\rho\eta)}{2}\right)\\\nonumber
                                         &\to&g_{in}^{*}(\eta) = \e^{i\omega_{in}\eta}.\nonumber
\end{eqnarray}
From this follow (\ref{pe38}) and (\ref{pe39}).

In a similar way we can construct the solutions for $\eta\to \infty$. By using the transformation
\begin{eqnarray}\label{pe40}
&&_{2}F_{1}(\epsilon, \upsilon;\varpi;z) = \frac{\Gamma(\varpi)\Gamma(\varpi-\epsilon-\upsilon)}{\Gamma(\varpi-\epsilon)\Gamma(\varpi-\upsilon)}\times_{2}F_{1}(\epsilon,\upsilon;\epsilon+\upsilon-\varpi;1-z)\\\nonumber
&&+(1-z)^{\varpi-\epsilon-\upsilon}\frac{\Gamma(\varpi)\Gamma(\epsilon+\upsilon-\varpi)}{\Gamma(\epsilon)\Gamma(\upsilon)}\times_{2}F_{1}(\varpi-\epsilon,\varpi-\upsilon;\varpi-\epsilon-\upsilon+1;1-z),\nonumber
\end{eqnarray}
where ${\Gamma}$ is the Euler function, identifying the coefficients 
\begin{eqnarray}\label{pe41}
\begin{cases}
\varpi-\epsilon-\upsilon = -\frac{i\omega_{out}}{\rho}\\
\varpi-\epsilon = \frac{1\pm(\sqrt{1-4r})^{*}}{2}-\frac{i\omega_{+}}{\rho}\\
\varpi-\upsilon = \frac{1\mp(\sqrt{1-4r})^{*}}{2}-\frac{i\omega_{+}}{\rho}\\
\epsilon+\upsilon-\varpi=\frac{i\omega_{out}}{\rho}\\
\varpi-\epsilon-\upsilon+1=1-\frac{i\omega_{out}}{\rho}\\
\end{cases}, 
\end{eqnarray}
for ${g^{(+)}_{in}(\eta)}$, and
\begin{eqnarray}\label{pe42}
\begin{cases}
\varpi-\epsilon-\upsilon = -\frac{i\omega_{out}}{\rho}\\
\varpi-\epsilon = \frac{1\pm(\sqrt{1-4r})^{*}}{2}-\frac{i\omega_{-}}{\rho}\\
\varpi-\upsilon = \frac{1\mp(\sqrt{1-4r})^{*}}{2}-\frac{i\omega_{-}}{\rho}\\
\epsilon+\upsilon-\varpi=\frac{i\omega_{out}}{\rho}\\
\varpi-\epsilon-\upsilon+1=1-\frac{i\omega_{out}}{\rho}\\
\end{cases}, 
\end{eqnarray}
for ${g^{(-)}_{in}(\eta)}$, we arrive at:
\begin{eqnarray}\label{pe43}
&&g_{in}^{(+)}(\eta) = \exp\left[i\left(-\omega_{+}\eta - \frac{\omega_{-}}{\rho}\ln(2\cosh(\rho\eta))\right)\right]\\\nonumber
&&\times\Bigg[{C_{1}}\;\times \;_{2}F_{1}\left(\frac{1\mp(\sqrt{1-4r})^{*}}{2} + \frac{i\omega_{-}}{\rho}, \frac{1\pm(\sqrt{1-4r})^{*}}{2} + \frac{i\omega_{-}}{\rho};\frac{i\omega_{out}}{\rho};\frac{1-\tanh(\rho\eta)}{2}\right)\\\nonumber
&&+{C_{2}}\;(1-z)^{-\frac{i\omega_{out}}{\rho}}\\\nonumber
&&\times \; _{2}F_{1}\left(\frac{1\pm(\sqrt{1-4r})^{*}}{2} - \frac{i\omega_{+}}{\rho}, \frac{1\mp(\sqrt{1-4r})^{*}}{2} - \frac{i\omega_{+}}{\rho};1-\frac{i\omega_{out}}{\rho};\frac{1-\tanh(\rho\eta)}{2}\right)\Bigg],\nonumber
\end{eqnarray}
\begin{eqnarray}\label{pe45}
&&g_{in}^{(-)}(\eta) = \exp\left[i\left(-\omega_{-}\eta - \frac{\omega_{+}}{\rho}\ln(2\cosh(\rho\eta))\right)\right]\\\nonumber
&&\times \Bigg[C_{3}\;\times \;_{2}F_{1}\left(\frac{1\mp(\sqrt{1-4r})^{*}}{2} + \frac{i\omega_{+}}{\rho}, \frac{1\pm(\sqrt{1-4r})^{*}}{2} + \frac{i\omega_{+}}{\rho};\frac{i\omega_{out}}{\rho};\frac{1-\tanh(\rho\eta)}{2}\right)\\\nonumber
&&+C_{4}\;(1-z)^{-\frac{i\omega_{out}}{\rho}}\\\nonumber
&& \times \; _{2}F_{1}\left(\frac{1\pm(\sqrt{1-4r})^{*}}{2} - \frac{i\omega_{-}}{\rho}, \frac{1\mp(\sqrt{1-4r})^{*}}{2} - \frac{i\omega_{-}}{\rho};1-\frac{i\omega_{out}}{\rho};\frac{1-\tanh(\rho\eta)}{2}\right)\Bigg],\nonumber
\end{eqnarray}
with
\begin{eqnarray}\label{pe44}
&& C_{1} = \frac{\Gamma(1-\frac{i\omega_{in}}{\rho})\Gamma(-\frac{i\omega_{out}}{\rho})}{\Gamma(\frac{1\pm(\sqrt{1-4r})^{*}}{2} - \frac{i\omega_{+}}{\rho})\Gamma(\frac{1\mp(\sqrt{1-4r})^{*}}{2} - \frac{i\omega_{+}}{\rho})},\\
&& C_{2} = \frac{\Gamma(1-\frac{i\omega_{in}}{\rho})\Gamma(\frac{i\omega_{out}}{\rho})}{\Gamma(\frac{1\mp(\sqrt{1-4r})^{*}}{2} + \frac{i\omega_{-}}{\rho})\Gamma(\frac{1\pm(\sqrt{1-4r})^{*}}{2} + \frac{i\omega_{-}}{\rho})}.
\end{eqnarray}
\begin{eqnarray}\label{pe46}
&& C_{3} = \frac{\Gamma(1+\frac{i\omega_{in}}{\rho})\Gamma(-\frac{i\omega_{out}}{\rho})}{\Gamma(\frac{1\pm(\sqrt{1-4r})^{*}}{2} - \frac{i\omega_{-}}{\rho})\Gamma(\frac{1\mp(\sqrt{1-4r})^{*}}{2} - \frac{i\omega_{-}}{\rho})},\\
&& C_{4} = \frac{\Gamma(1+\frac{i\omega_{in}}{\rho})\Gamma(\frac{i\omega_{out}}{\rho})}{\Gamma(\frac{1\mp(\sqrt{1-4r})^{*}}{2} + \frac{i\omega_{+}}{\rho})\Gamma(\frac{1\pm(\sqrt{1-4r})^{*}}{2} + \frac{i\omega_{+}}{\rho})}. 
\end{eqnarray}
By using (\ref{pe31}) and 
\begin{equation}\label{pe48}
\frac{1+tgh(\rho\eta)}{2} = \frac{e^{2\rho\eta}}{e^{2\rho\eta}+1}, 
\end{equation}
Eqs. (\ref{pe43}) and (\ref{pe45}) are rewrite as:
\begin{eqnarray}\label{pe49}
&& g_{in}^{(+)}(\eta) = \exp\left[i\left(-\omega_{+} \eta - \frac{\omega_{-}}{\rho}\ln(2\cosh(\rho\eta))\right)\right]\\\nonumber
&&\times C_{1}\times \; _{2}F_{1}\left(\frac{1\mp(\sqrt{1-4r})^{*}}{2} + \frac{i\omega_{-}}{\rho},\frac{1\pm(\sqrt{1-4r})^{*}}{2} + \frac{i\omega_{-}}{\rho};\frac{i\omega_{out}}{\rho};\frac{1-tgh(\rho\eta)}{2}\right)\\\nonumber
&&+C_{2}\times \exp\left[i\left(-\omega_{+}\eta - \frac{\omega_{-}}{\rho}\ln(2\cosh(\rho\eta))\right)\right]\times \left(\frac{\e^{2\rho\eta}}{\e^{2\rho\eta}+1}\right)^{\frac{i\omega_{in}}{\rho}}\times\left(\frac{1}{\e^{2\rho\eta}+1}\right)^{-\frac{i\omega_{out}}{\rho}}\\\nonumber
&&\times\;_{2}F_{1}\left(\frac{1\mp(\sqrt{1-4r})^{*}}{2} - \frac{i\omega_{-}}{\rho},\frac{1\pm(\sqrt{1-4r})^{*}}{2} - \frac{i\omega_{-}}{\rho};1-\frac{i\omega_{out}}{\rho};\frac{1-\tanh(\rho\eta)}{2}\right)\nonumber
\end{eqnarray}
and
\begin{eqnarray}\label{pe50}
&&g_{in}^{(-)}(\eta) = \exp\left[i\left(-\omega_{-}\eta - \frac{\omega_{+}}{\rho}\ln(2\cosh(\rho\eta))\right)\right]\\\nonumber
&&\times C_{3}\times\; _{2}F_{1}\left(\frac{1\mp(\sqrt{1-4r})^{*}}{2} + \frac{i\omega_{+}}{\rho},\frac{1\pm(\sqrt{1-4r})^{*}}{2} + \frac{i\omega_{+}}{\rho};\frac{i\omega_{out}}{\rho};\frac{1-\tanh(\rho\eta)}{2}\right)\\\nonumber
&&+C_{4}\times \exp\left[i\left(-\omega_{-}\eta - \frac{\omega_{+}}{\rho}\ln(2\cosh(\rho\eta))\right)\right]\times \left(\frac{\e^{2\rho\eta}}{\e^{2\rho\eta}+1}\right)^{-\frac{i\omega_{in}}{\rho}}\times\left(\frac{1}{\e^{2\rho\eta}+1}\right)^{-\frac{i\omega_{out}}{\rho}}\\\nonumber
&&\times_{2}F_{1}\left(\frac{1\mp(\sqrt{1-4r})^{*}}{2} - \frac{i\omega_{+}}{\rho},\frac{1\pm(\sqrt{1-4r})^{*}}{2} - \frac{i\omega_{+}}{\rho};1-\frac{i\omega_{out}}{\rho};\frac{1-\tanh(\rho\eta)}{2}\right).\nonumber
\end{eqnarray}
Finally, we can identify ${g_{out}^{(+)}(\eta)}$, ${g_{out}^{(-)}(\eta)}$, ${\sigma}$, ${\varrho}$, ${\pi}$ and ${\tau}$ as in (\ref{pe53})-(\ref{pe54}) and the coefficients (\ref{pe55})-(\ref{pe58}).

\begin{acknowledgements}
SHP is grateful to CNPq - Conselho Nacional de Desenvolvimento Cient\'ifico e Tecnol\'ogico, Brazilian research agency, for financial support, grants number 304297/2015-1. RCL is grateful to CAPES for financial support.
\end{acknowledgements}

\newpage


\end{document}